\documentclass[structabstract]{aa}  
\usepackage{graphicx}
\usepackage[varg]{txfonts}
\usepackage{natbib}
\bibpunct{(}{)}{;}{a}{}{,} 

\newcommand{\cgs}{$10^{-9}\ \mathrm{ergs}\ \mathrm{cm}^{-2}\ \mathrm{s}^{-1}\ \mathrm{sr}^{-1}\ \AA^{-1}$\ }
\newcommand{\cgss}{$\mathrm{ergs}\ \mathrm{cm}^{-2}\ \mathrm{s}^{-1}\ 
\mathrm{sr}^{-1}$\ }

\begin{document}

   \title{Spectroscopy of diffuse light in dust clouds}

   \subtitle{Scattered light and the solar neighbourhood radiation field
             \thanks{Based on observations collected at the European 
             Southern Observatory, Chile, under programme ESO No.\
             073.C-0239(A)}}

   \author{K.\ Lehtinen  \inst{1,2} 
           \and 
           K.\ Mattila   \inst{2} }

   \institute{Department of Physics, P.O.B.\ 64, University of Helsinki, FI-00014 Helsinki\\ \email{kimmo.lehtinen@helsinki.fi}
\and 
Observatory\thanks {Closed down on December 31, 2009}, University of Helsinki, FI-00014 Helsinki \\ 
            \email{kalevi.mattila@helsinki.fi}     }

   \date{Accepted for publication in Astronomy \& Astrophysics }

 
  \abstract
  {The optical surface brightness of dark nebulae is mainly due to  scattering of integrated starlight by classical dust grains. It contains information on the impinging interstellar radiation field, cloud structure, and grain scattering properties. We have obtained spectra of the scattered light from 3500 to 9000~\AA\ in two globules, the Thumbprint Nebula and DC~303.8-14.2.}
  {We use observations of the scattered light to study the impinging integrated starlight spectrum as well as the scattered H$\alpha$ and other line emissions from all over the sky.  We search also for the presence of other than scattered light in the two globules.}
{We obtained long-slit spectra encompassing the whole globule  plus adjacent sky in a one-slit setting, thus enabling efficient elimination of airglow and other foreground sky components. We calculated synthetic integrated starlight spectra for the solar neighbourhood using HIPPARCOS-based stellar distributions and the spectral library of Pickles.}
{Spectra are presented separately for the bright rims and dark cores of the globules. The continuum spectral energy distributions and absorption line spectra can be well modelled with the synthetic integrated starlight spectra. Emission lines of H$\alpha$+[\ion{N}{ii}], H$\beta$, and [\ion{S}{ii}] are detected and are interpreted in terms of scattered light plus an \mbox{{\em in situ}} warm ionized medium component behind the globules. We detected an excess of emission over the wavelength range 5200--8000~\AA\ in DC~303.8-14.2 but the nature of this emission remains open.} 
   {}
 

   \maketitle
%

\section{Introduction}

Since the pioneering work by \citet{s36}, the surface brightness of dark nebulae has been recognised as a tool for studying the scattering properties of interstellar grains. Broadband photometric studies have been used to constrain the grain albedo and scattering  phase function in the optical 
\citep{m70a,f76}, near-IR \citep{l96} and UV \citep{h95}. Because of the faint surface brightness levels involved, intermediate-resolution photometric or spectrophotometric studies have been carried out only in a few cases so far \citep{m79,lms87}. In these studies the impinging interstellar radiation field (ISRF) has been  assumed to be known in the solar neighbourhood. Broad band observational determinations of the optical ISRF include \citet{eh60} and \citet{t81}, while some frequently used ISRF models  have been published by \citet{h68} and \citet{mmp83}. A higher resolution spectral energy distribution(SED) and line spectrum of the optical ISRF has been calculated  by \citet{m80a,m80b} using population synthesis modelling of the integrated starlight.

Higher spectral resolution observations of the surface brightness of dark nebulae are desirable for at least three reasons: (i) to study the SED of the ISRF in more detail with the aim of justifying the population synthesis modelling; (ii) to study the wavelength dependence of the dust-scattering properties in greater detail; and (iii) to find out whether other mechanisms, such as photoluminescence, contribute to the surface brightnes in addition to scattering. In this paper we are concerned with items (i) and (iii).   

The observed optical/near-IR scattered light spectrum of a dust cloud, externally illuminated by the Galactic ISRF, can be used to empirically  determine the spectrum of the ISRF. Indeed, the clouds can be used as a kind of ISRF remote sensing measuring devices or modified Ulbricht spheres 
\citep{u20}, placed in different locations of the solar neighbourhood. Recently, \citet{m11} have published the spectrum 
(3500--6000~\AA) of the high galactic latitude dark cloud \object{L1642} and found a good agreement with the integrated starlight spectrum synthesized for the solar neighbourhood. \citet{bd12} have determined the average  scattered light spectrum of diffuse dust by using 92000 blank-sky-spectra from the Sloan Digital Sky Survey. 
  
Spectroscopic observations are also needed to check whether the extended red emission (ERE), detected so far in reflection nebulae, planetary nebulae, HII regions, galaxies, and the diffuse interstellar medium is present also in dark nebulae.  A previous spectrophotometric measurement of the dark nebula L 1780 by \citet{m79} has been interpreted by \citet{cl87} to show evidence for ERE. Because of the low spectral resolution (R $\sim$ 30) and relatively narrow wavelength coverage (5000 to 7500~\AA) of this measurement, there is obvious need for additional, wider band and higher resolution spectral measurements of the surface brightness of dark nebulae. This need was emphasised also in the compilation and comprehensive discussion of the existing ERE data by 
\citet{sw02}, where \object{L1780} had an important place.

Based on observations made with the ESO Multi-Mode Instrument (EMMI) 
\citep{dekker86} at the 3.5-m NTT at ESO La Silla Observatory, we present spectra from 3500 to  9000~\AA\ of the scattered light in two globules. Our observations extend the number of dark clouds observed for scattered light spectrum from one to three and the wavelength coverage up to 9000~\AA.

The two globules of this study, the Thumbprint Nebula (\object{DC302.6-15.9}) and \object{DC303.8-14.2} (to be called TPN and DC~303 for brief), are located in the Chamaeleon III molecular  cloud area. For the description of the globules and a review of previous optical, infrared, and radio molecular line observations see \citet{f74}, \citet{f76}, \citet{l95}, \citet{l96}, \citet{llm98} and \citet{k07}. The TPN has a regular elliptical shape with a bright rim - dark core structure. Its apparent size is $\sim 5 \times 3$ arcmin. DC~303 is somewhat larger, $\sim 10 \times 5$ arcmin, less regular in shape but with a similarly pronounced bright rim - dark core structure as TPN. The density structure of TPN and DC~303 has been studied using deep {\it JHK} photometry \citep{k07}. DC~303 has a dense centrally peaked core with extinction $A_V>50$~mag, while the TPN has a more transparent core with a peak extinction of
$A_V \approx 13$~mag. DC~303 has an optically invisible embedded far-infrared source near its centre, a Class 0/I young stellar object. The TPN is starless \citep{k07}. As a distance estimate for both globules we use the value for Cha~III, 217$\pm12$~pc \citep{k10}.
  
The surface brightnesses in this paper is given in units of \cgs, for which we use the abbreviation cgs. For the integrated spectral line intensities we use the abbreviation 
1~cgs\AA=$10^{-9}\ \mathrm{ergs}\ \mathrm{cm}^{-2}\ \mathrm{s}^{-1}\ 
\mathrm{sr}^{-1}$.


\section{Observations}

\begin{table*}
\caption{Instrument setup of the EMMI instrument for spectroscopic observations. The spectral resolution, measured from arc-lamp spectra, is defined in Sect.~3.3. The wavelength range given is the range used for making the spectrum. }  
\centering 
\begin{tabular}{c c c c c} 
\hline\hline
Mode & Dispersive & Dispersion\tablefootmark{a}  & Spectral & Wavelength  \\ 
     & element    & (\AA/pixel) & resolution (\AA) & range (\AA)    \\
\hline 
BLMD\tablefootmark{b} & grating \#4 & 3.68 & 44 & 3381--5119       \\ 
RILD\tablefootmark{c} & grism \#3   & 2.86 & 25 & 4302--9012 \\
\hline 	
\end{tabular}
\tablefoot{
\tablefoottext{a}{For $2\times2$ pixel binning used in the observations}
\tablefoottext{b}{Blue medium dispersion spectroscopy} 
\tablefoottext{c}{Red imaging and low dispersion spectroscopy} }
\label{tab:InstrumSetup}
\end{table*}

\begin{figure}
\centering
\resizebox{\hsize}{!}{\includegraphics{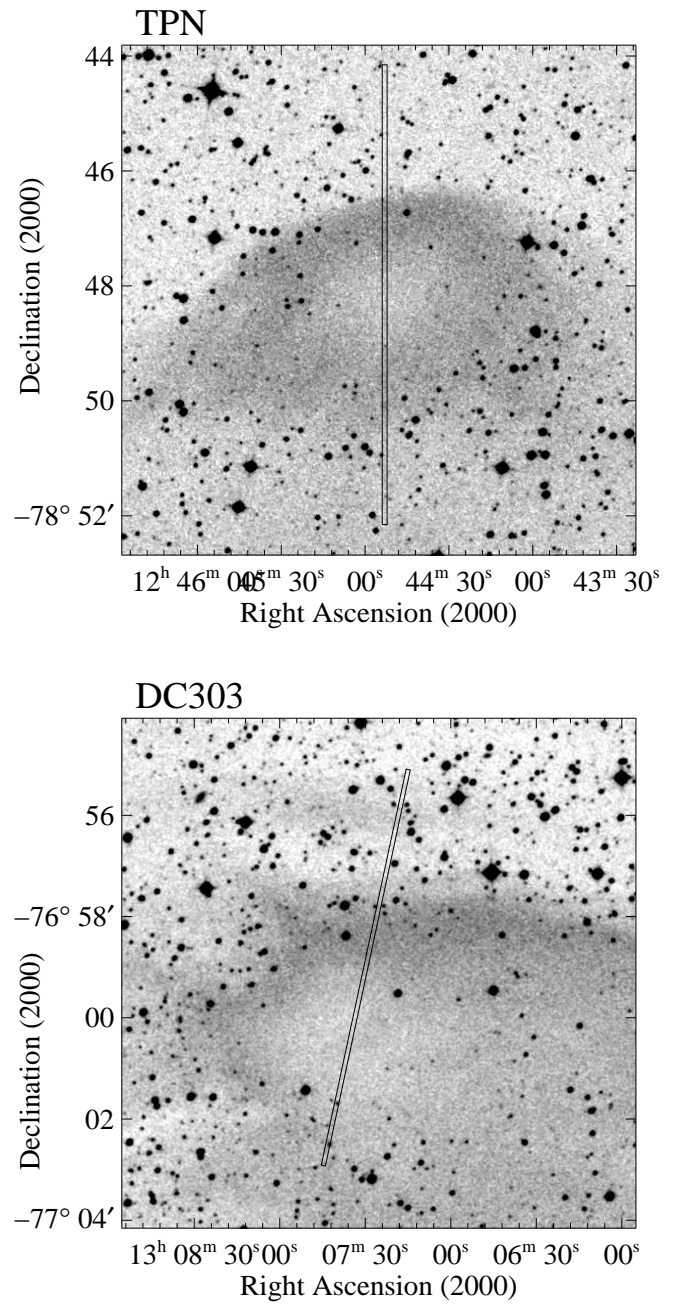}}
\caption{Slit positions superimposed on Digitized Sky Survey red images of the TPN and DC~303. The length of the slit, 8\arcmin, is that of the slit in the red arm of the spectrometer. The position angle (East of North) of the slit is 
0\degr\ and 168\degr\ for TPN and DC~303, respectively.}
\label{fig:DSS-images}
\end{figure}

The observations were made with the EMMI \citep{dekker86} at the New Technology Telescope (NTT) on the La Silla observatory during the nights of 17--19 May 2004. The spectroscopic  observing modes used were BLMD (blue medium dispersion spectroscopy) in the blue arm of the instrument and  RILD (red imaging and low dispersion spectroscopy) in the red arm.  The detector in the red arm consists of two CCD chips (2048$\times$4096 pixels) with a gap of 7.5\arcsec\. We used $2\times2$ binning, giving a pixel scale of 0.33\arcsec/pixel. The detector in the blue arm is a single 1024$\times$1024 pixel chip that was binned by $2\times2$ pixels, giving a scale of 0.72\arcsec/pixel. In the BLMD mode we used grating \#4 as the dispersive element. In the RILD mode we used grism \#3 as the dispersive element. The slit width was set to 5\arcsec. The slit length was $\sim$6\arcmin\, and $\sim$8\arcmin\, in the blue and red arm of EMMI, respectively. Details of the instrument setup are listed in Table~\ref{tab:InstrumSetup}. The integration time of a single spectrum is 30 minutes. For the intensity calibration of spectra we measured several spectrophotometric standard stars each night through a 5\arcsec\ slit. The wavelength calibration of spectra is based on arc lamp frames. 

The slit of EMMI was positioned through the centre of the rather spherical clouds in such a way that the slit included the dark core, bright rim(s), and pure background sky (see Fig.~\ref{fig:DSS-images}). The diameter of TPN, 
$\sim$5\arcmin, is such that both of the bright rims fit inside the slit. Because the diameter of DC~303 is larger than the length of the slit, we measured only one bright rim, the dark core, and the background sky. Digitized sky survey images were used to set the centre position and orientation angle of the slit in a way that minimises the number of stars within the slit.  

In this project, we made spectroscopic long-slit observations of the weak surface brightness of a dust cloud that extends over a considerable length of the slit. Furthermore, we used blank sky areas in the science frames to determine the spectrum of the night sky emission, which was then subtracted from the object spectrum. For a proper calibration of these data, two items require special attention. Firstly, we must be able to make a good flat-field correction for the large-scale gradients over the frame. The best flat-field correction, both for imaging and spectroscopy, is obtained when the illumination of the instrument and the spectrum of the radiation entering the instrument are the same between flat-field calibration images and science images. We fulfilled these conditions by taking separate flat-field images during the nighttime, pointing the telescope to a position near the globules that is free of gradients in dust emission, based on 100\,$\mu$m IRAS images. Secondly, we must be able to remove the geometrical distortions in the dispersion axis of the spectra, to ensure that the dispersion axis is along a row or column over the whole frame. Otherwise, subtracting the night sky emission spectrum produces artefacts on the object spectrum, especially at wavelengths where the gradient in the night sky spectrum is at steepest. We performed this geometrical correction by using the arc lamp calibration spectra of the EMMI instrument. 


\section{Data reduction}

Data reduction was done using IRAF\footnote{IRAF is distributed by the National Optical Astronomy Observatories, which are operated by the Association of Universities for Research in Astronomy, Inc., under cooperative agreement with the National Science Foundation} (Image Reduction and Analysis Facility) \citep{tody93}. The details of the data reduction are the following. \\ 
i) Bias subtraction. We used separate bias images to determine the bias level in the science frames. At the blue arm the master bias image is flat without structure, so the bias value was taken to be a scalar and subtracted. At the red arm the master bias image shows a gradient across the frame, therefore a 2D bias image was subtracted. \\
ii) Flat-field correction. To make the flat-field correction for the large-scale gradients over the images, we used the nightsky flat-fields as discussed in Sect.~2. These flat-field images are weakly illuminated and therefore do not have a sufficiently high signal-to-noise ratio. Therefore, we first replaced every line of the flat field image in the spatial direction with a Chebyshev function fit of the line, and discarded any stars in the fit. Each line was then scaled to a mean value of unity. We divided the science frames with each individual flat-field image and used the flatness of the background sky outside the globules to decide which flat field image gives the best correction. To flat-field-correct for pixel-to-pixel sensitivity variations we used the dome flat images. We first took a mean of all dome flat images, then smoothed the mean image with a Gaussian function, and then divided the mean image with its smoothed version. This provided a flat field image that only includes the pixel-to-pixel sensitivity variations.

We corrected for geometrical distortion of the spectra in the following way in IRAF: i) on a 2D arc-lam spectrum the emission features along a single dispersion line are identified with the \texttt{identify} task of IRAF;
ii) the emission features at other dispersion lines are re-identified with the 
\texttt{reidentify} task; iii) the wavelengths of the identified features as a function of pixel position are fitted with a two-dimensional function using the \texttt{fitcoords} task; iv) the geometrical correction is made with the \texttt{transform} task, after which the wavelength is a linear function along one axis (dispersion is constant). Despite this correction, subtracting the background sky emission produces artefacts on spectra at the edges of the brightest airglow emission line profiles where the intensity gradient is  steepest. Therefore, we interpolated the following wavelength ranges of the spectra:  
5544~\AA--5601~\AA, 5909~\AA--5918~\AA, and 6277~\AA--6322~\AA.

\subsection{Atmospheric extinction}

\begin{figure}
\resizebox{\hsize}{!}{\includegraphics{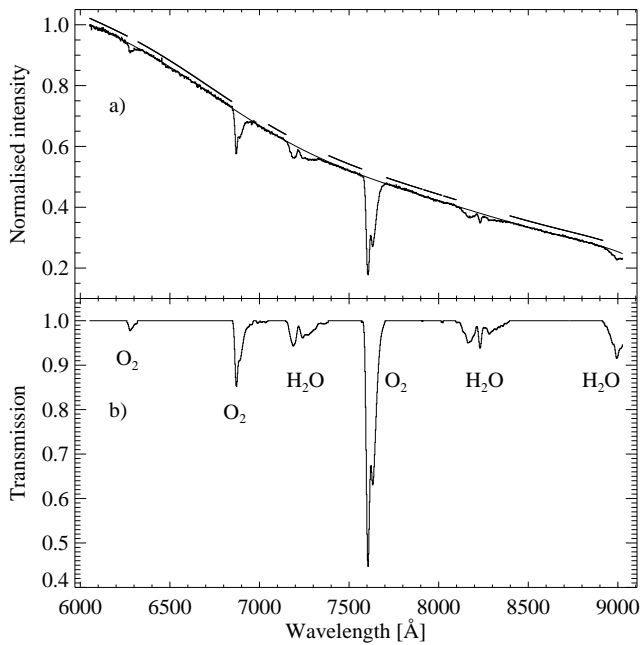}}
\caption{ Derivation of atmospheric transmission. The observed spectrum of the
standard star LTT4362 (histogram), overplotted by a Chebyshev fit (solid line)
to the continuum of the spectrum. The regions that are defined to
represent the continuum are shown as intermittent lines above the fit (a).
The derived atmospheric transmission spectrum at airmass of unity (b). 
The molecular bands of the spectrum are identified. }
\label{fig:transmission}
\end{figure}

The observed spectra have to be corrected for atmospheric absorption lines. To derive a spectrum of telluric absorption we used the spectrum of the standard star LTT4364 observed by us. The spectral class of LTT4364 is DQ6.4 \citep{s09} and it is free of significant intrinsic absorption features in the wavelength range 6000--9000~\AA. At the spectral resolution of our observations, about 44~\AA, the individual, narrow absorption lines blend into wide absorption bands that have wide and weak wings (see Fig.~\ref{fig:transmission}). Thus it is difficult to determine where the continuum (areas not affected by telluric absorption) in the spectrum of LTT4364 is located. To determine the wavelength ranges of the continuum we used the tabulated atmospheric transmission spectrum published by NOAO (National Optical Astronomy Observatories) (ftp://ftp.noao.edu/catalogs/atmospheric\_transmission). Firstly, we convolved the NOAO spectrum to the resolution of our observations. A comparison between the convolved NOAO spectrum and the LTT4364 spectrum verifies that the absorption features seen in the LTT4364 spectrum are all caused by telluric absorption. Secondly, we defined the continuum as wavelength ranges where atmospheric transmission in the convolved  NOAO spectrum is higher than 99\%. Then, we fitted the observed spectrum of LTT4362 with a Chebyshev polynomial, excluding the data points that are outside the continuum, as defined above.
The observed spectrum of LTT4364 and the continuum fit are shown in 
Fig.~\ref{fig:transmission}a. By dividing the spectrum of LTT4364 with the fit, we obtain the normalised telluric transmission. As the final steps, we i) applied a Savitzky-Golay smoothing filter to the derived transmission spectrum (reducing noise while retaining the dynamic range of the variations); ii) defined the transmission to have a value of unity in the continuum regions as defined above; and iii) scaled the transmission to transmission at airmass of unity by raising the transmission to a power of (1/$X$) where $X$ is the airmass of LTT4364 during our observation. The final transmission spectrum is shown in Fig.~\ref{fig:transmission}b. At wavelengths longer than about 8900~\AA\ the derived transmission is extrapolation because of the absence of a continuum region at the long wavelength side of the absorption feature.

Each individual globule spectrum was corrected for telluric absorption by dividing by the transmission spectrum (Fig.~\ref{fig:transmission}b) raised to the power of $X$, $X$ being the airmass of the globule. Correcting the deep absorption feature at $\sim 7600$~\AA\ leaves noticeable residuals in the spectra, therefore we interpolated the wavelength range 
7568~\AA--7652~\AA\ after the telluric absorption correction. 

\subsection{Extracting the spectra}
The main products to be derived from our spectroscopic data are the spectra of scattered light at the position of the bright rims and the dark cores. Because the optical depth through a dust cloud depends on the wavelength, the shape of the surface brightness profile changes accordingly (see 
Fig.~\ref{fig:2Dspectrum}). Thus, one should determine the extent of the bright rim and the dark core in the spatial direction in a way that is dependent of the wavelength. 

In the red arm we used the following method. 
Firstly, to obtain a sufficient signal-to-noise ratio, we rebinned the surface brightness profiles over wavelength ranges of $\sim$33~\AA\ (11 pixels). The rebinned profiles were fitted with a Chebyshev function in spatial direction, and the pixel locations of the maxima of the two bright rims and the minimum of the dark core were noted. The bright rim was defined as a window over those pixels in spatial direction where the surface brightness exceeds the maximum of the bright rim by 0.8 times, and the dark core was defined as a window where the surface brightness is less than 1.2 times the minimum of the dark core. The derived extents of the rims and the core as a function of wavelength were fitted with a low-order polynomial, and the fit was used to interpolate the windows at every wavelength of the full-resolution spectra. The windows of the bright rims and dark core defined in this way are shown in 
Fig.~\ref{fig:2Dspectrum} for a single spectrum of TPN. 

In the blue arm the signal-to-noise ratio in individual spectra is so low that this method is not applicable. Therefore, we used constant ranges in spatial direction to specify the extent of the bright rim and the dark core.

We then took the mean of the pixel values at each wavelength over the spatial  windows as defined above. The blank sky at the northern side of the globules was used to determine the intensity level of the background sky at each wavelength, which was then subtracted. We defined the intensity $\Delta$I as an intensity of diffuse light of the globule relative to the intensity of the background sky. To improve the signal-to-noise ratio, we rebinned the spectra into 20~\AA\ wide bins.

\subsection{Wavelength resolution}
The point spread function (PSF) of the spectra in dispersion direction resembles a square-box profile, as determined from emission line profiles in arc-lamp measurements. Therefore we give the wavelength resolution as effective width, determined as the area under an emission line profile divided by the maximum of the profile. In the red arm, the resolution changes from about 42~\AA\ to about 45~\AA\ from the short- to the long-wavelength edge of the spectra.  The mean value of the resolution over 13 separate  emission lines of an arc-lamp spectrum is 44$\pm$1~\AA. In the blue arm, the resolution over four separate  emission lines is 25$\pm$1~\AA, with no dependence on wavelength. 

\subsection{Error analysis}
Most of the total signal measured by us is nightsky emission. Consequently, the total number of detected photons varies strongly as a function of wavelength (see Fig.~\ref{fig:2Dspectrum}). The maximum surface brightness of the diffuse light from the globules is $\approx$1/10 of the brightness of the pseudo-continuum of nightsky emission (that is, outside the emission peaks). In the red arm of EMMI, the total number of detected electrons in a single exposure varies between $\sim$300--2500~e$^{-}$ per pixel, depending on the strength of atmospheric emission. For the pseudo-continuum the intensity is typically about 400~e$^{-}$ per pixel. The signal-to-noise ratio for diffuse light at the bright rims is thus about two per pixel on a single exposure.  When combining all individual exposures, we have typically $\sim$2000 pixels on spatial direction at each wavelength over which to calculate the mean value of the diffuse light intensity at the bright rims. The standard deviation of the diffuse light spectrum of the bright rim caused by photon noise is thus 
$\sim$1\% of the intensity of scattered light. This deviation can be considered negligible compared to other noise sources. The other sources of error that we considered are related to photometric calibration and imperfect flat-field correction and background subtraction, producing an estimated total uncertainty of 25\%.


\section{Results}

\begin{figure}
\centering
\resizebox{\hsize}{!}{\includegraphics{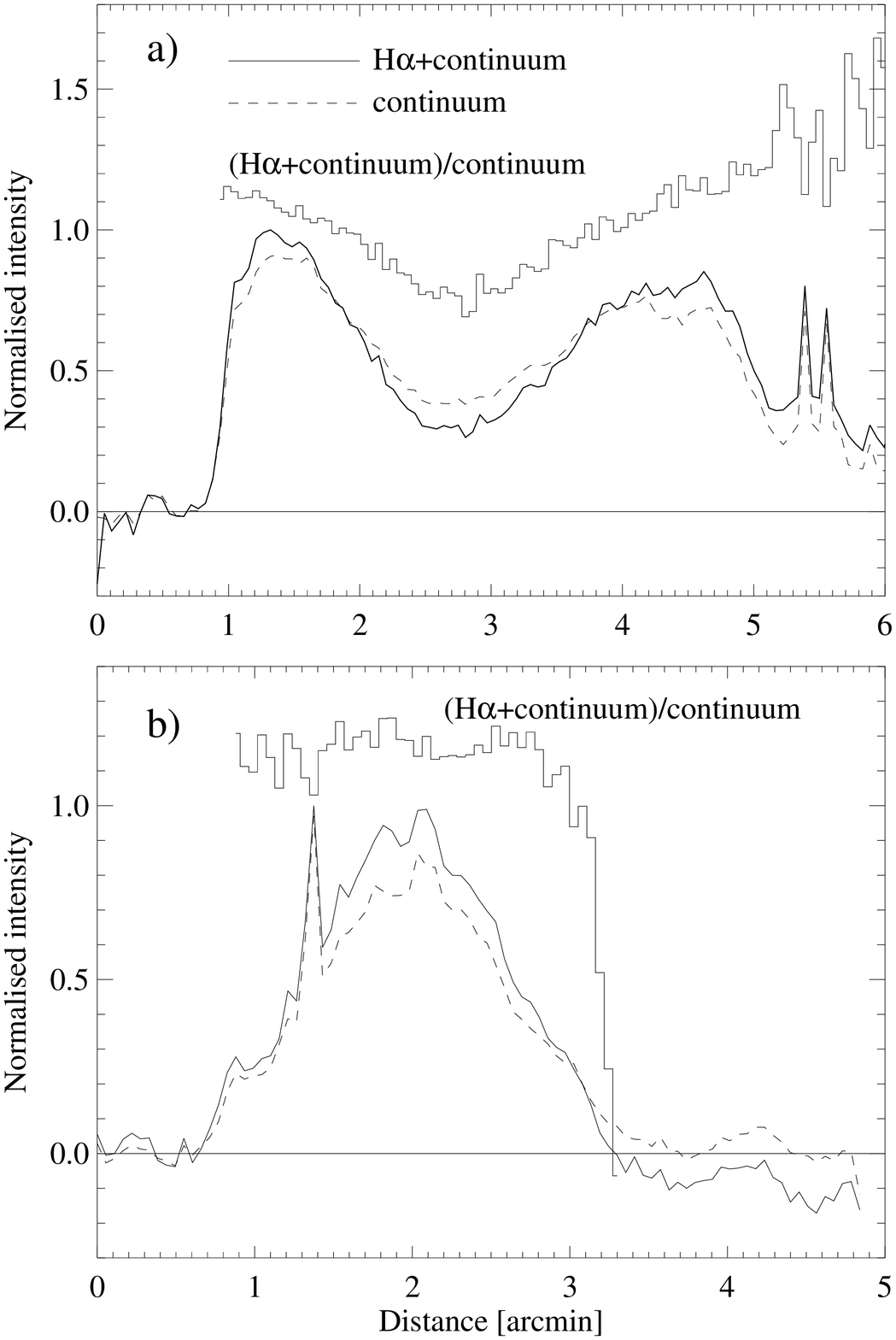}}
\caption{Surface brightness profiles across the globules in a band containing the H$\alpha$ and [\ion{N}{ii}] lines, together with the underlying continuum (solid line) and the adjacent continuum (dashed line) as well as their ratio (histogram). Panel a is for TPN, panel b for DC~303. The narrow peaks in the profiles are stars. The ratio is shown only for the region where the continuum intensity is not close to zero. The outer, northern boundary of DC~303 is on the left-hand side of the bright rim at $\sim$0.7\arcmin.} 
\label{fig:profiles}
\end{figure}

\begin{figure*}
\centering
\includegraphics{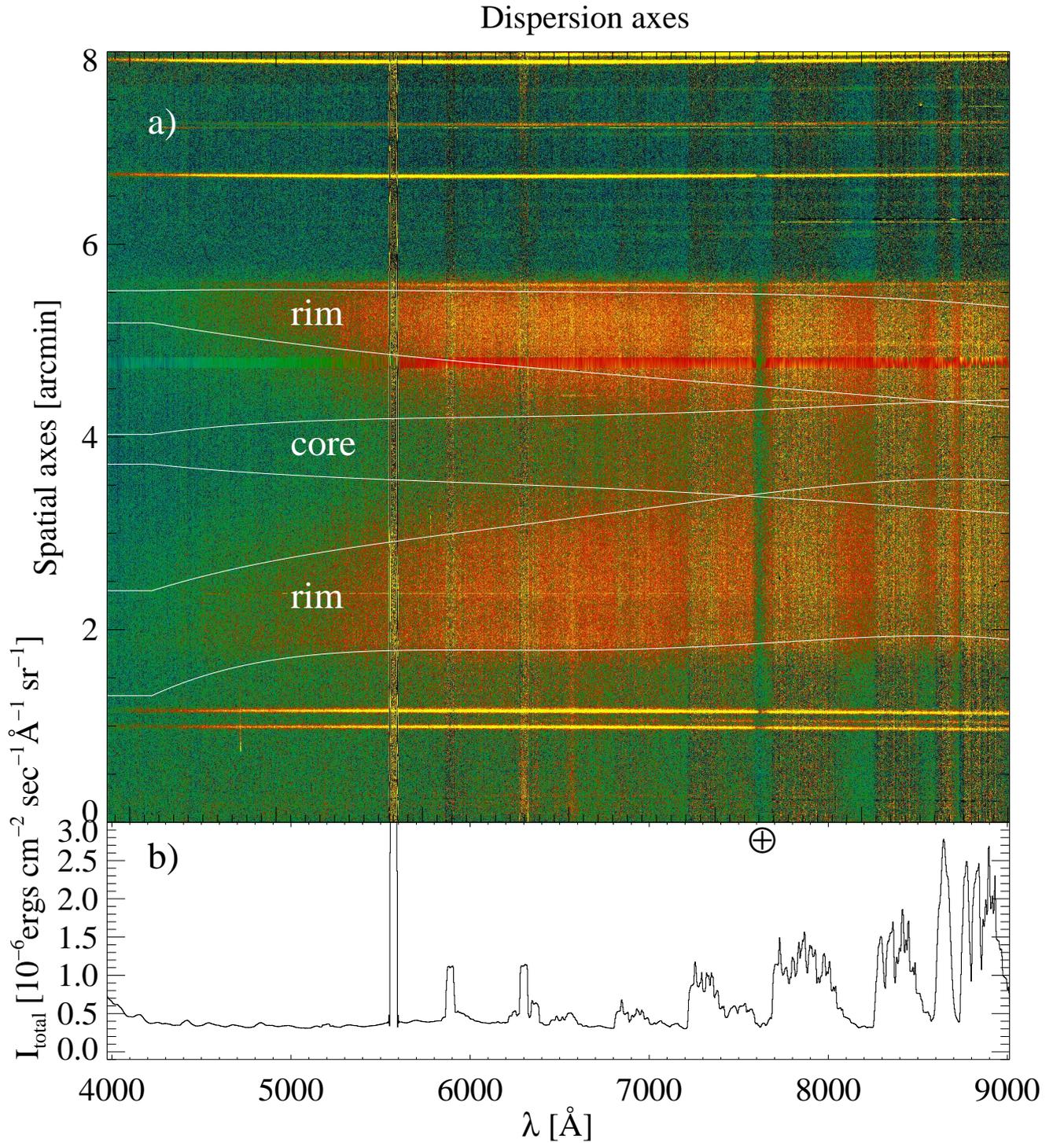}
\caption{ Example of a single spectrum of TPN as measured in the red arm  of EMMI. Panel a shows the spectrum of diffuse light after background subtraction. The bright rim--dark core structure of the globule is evident.  Panel b shows the total measured signal, i.e.\ the measured spectrum before the background subtraction when the dominating component is atmospheric emission.  Note the positive correlation between the noise of the scattered light and the strength of the atmospheric emission. The extents of the bright rims and the dark core, as defined in Sect.~3.2, are marked with white lines.  The location of the deep telluric absorption line at $\sim$7650~\AA\ is marked with a $\oplus$-sign. The horizontal region located at 
$\sim$4.3\arcmin\ on the spatial axes is the gap between the two CCD arrays. The gap is interpolated along columns (spatial axes). The intensity of the 5577~\AA\ [OI] atmospheric emission line is off-scale. Some stars outside the globule are seen as yellow strikes.  The sky north of TPN (upper side of panel a) is dimmer than the sky south of TPN (lower side in panel a) because of the cometary shape of the cloud. } 
\label{fig:2Dspectrum}
\end{figure*}

\begin{figure*}
\centering
\includegraphics[width=12cm]{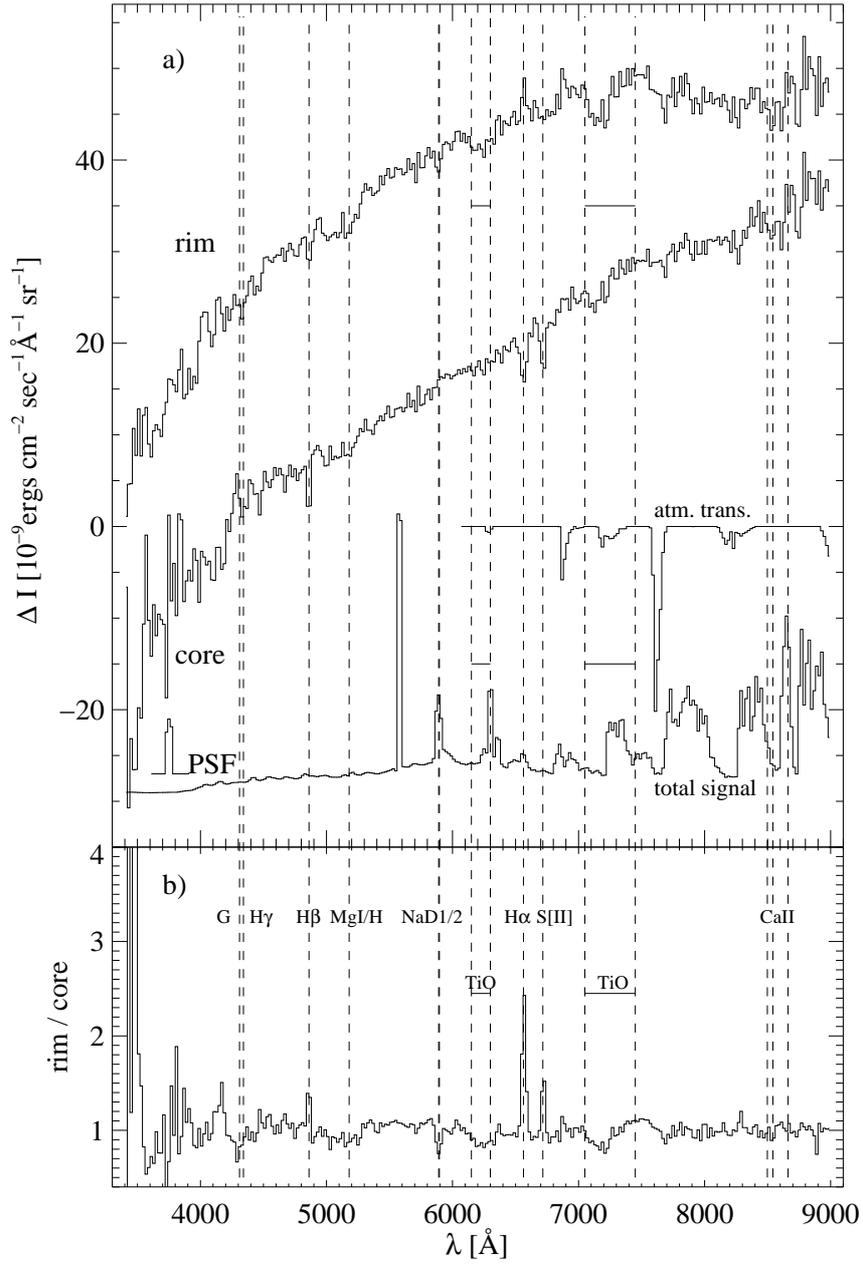}
\caption{Spectrum of diffuse light for TPN. Intensity $\Delta I$ is given  relative to the background sky adjacent to the globule.  Panel a shows the spectra for the dark core and the mean of the bright rims. Panel b shows the ratio of the rim and the core spectrum after fitting and subtracting the continuum of the spectra and scaling the spectra to unity. The wavelengths of spectral lines and bands are marked with dashed lines. The spectral resolution of the measurements (PSF) is shown. The atmospheric transmission between $\sim 6000-9000$~\AA\ and the total measured signal (including the atmospheric emission lines) are shown in panel a for reference, in arbitrary scale. }
\label{fig:TPNspectrum}
\end{figure*}

\begin{figure*}
\centering
\includegraphics[width=12cm]{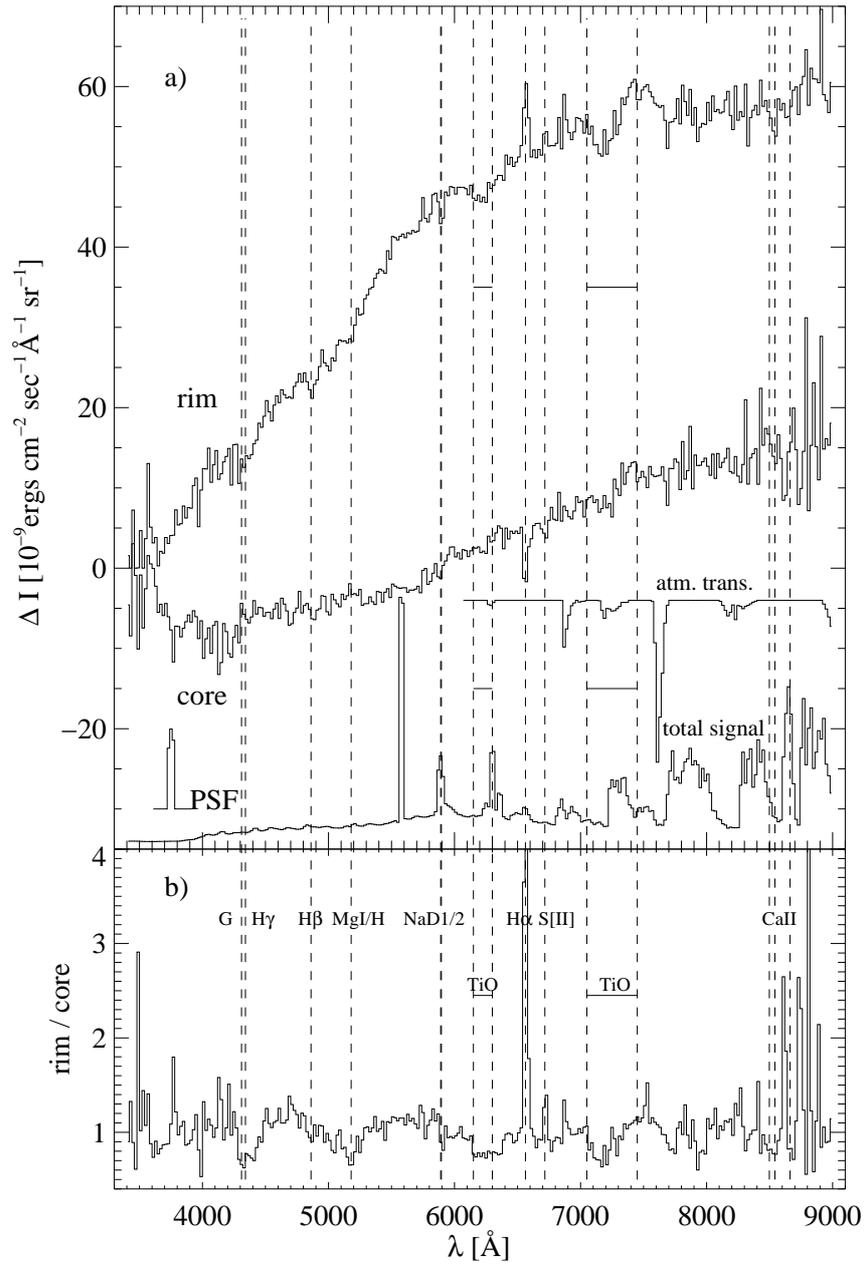}
\caption{Spectrum of scattered light for DC~303.8-14.2. The details are the same as for Fig.~\ref{fig:TPNspectrum} }
\label{fig:DC303spectrum}
\end{figure*}

\begin{table*}
\caption{ Detected and searched lines and bands in the spectra of scattered light. Results are given separately for rim and core spectrum. Columns 3--6: line is seen in emission ('em') or absorption ('abs') relative to the background sky, or not detected at all (blank). Columns 7 and 8: strong atmospheric emission or telluric absorption make the status of the line uncertain. }  
\centering 
\begin{tabular}{c c c c c c c c} 
\hline\hline
Line or & Wavelength & TPN rim & TPN core & DC~303 rim & DC~303 core & Atmosph. emission & Atmosph. absorption   \\ 
band & [\AA] & & & & & contamination & contamination  \\
1 & 2 & 3 & 4 & 5 & 6 & 7 & 8  \\
\hline
G-band+H$\gamma$  & 4300+4341 & abs & abs & abs &     &         & \\
H$\beta$          & 4861      & abs & abs & abs &     &         & \\
MgI + MgH         & 5180      & abs & abs & abs &     &         & \\
Na D1             & 5896      & abs &     & abs & abs & $\surd$ & \\
Na D2             & 5890      & abs &     & abs & abs & $\surd$ & \\
TiO               & [7050-7450] & abs &     & abs &     & $\surd$ & \\
H$\alpha$+[\ion{N}{ii}] & 6563+6548/6583 & em  & abs & em  & abs &  & \\
$[\ion{S}{ii}]$         & 6716/6731   &     & abs & em  & abs &         & \\
TiO               & [6150-6300] & abs & abs & abs & abs & $\surd$ & $\surd$ \\
\ion{Ca}{ii}      & 8498+8542   & abs & abs & abs & abs &         & \\
\ion{Ca}{ii}      & 8662        &     &     &     &     & $\surd$ & \\
\hline 	
\end{tabular}
\label{tab:LineDetection}
\end{table*}

\subsection{Surface brightness profiles}

Figure~\ref{fig:profiles} shows examples of the surface brightness profiles for TPN and DC~303 as measured along the slit, after subtracting the background sky emission. The region at the leftmost part of the plots represents background sky which has been used to determine the zero level of intensity at each wavelength. The profiles are for a band containing the H$\alpha$ and [\ion{N}{ii}] lines, together with the underlying continuum, and for an adjacent continuum band (mean value over the wavelength ranges 6473--6534~\AA\ and 6615--6676~\AA). Also shown is the ratio of these two bands. The surface brightness profiles were scaled to the maximum intensity of the band containing the H$\alpha$ and [\ion{N}{ii}] lines.

\subsection{Spectra of the bright rims and dark cores}
The spectra of scattered light are shown in Figs.~\ref{fig:TPNspectrum} and 
\ref{fig:DC303spectrum} for TPN and DC~303. Panels a show the spectra for rim and core, while panels b show the ratio of rim to core.  In the following we give a description of the observed properties of the lines and bands that were detected or searched for in our spectra.\\ 
H$\alpha$+[\ion{N}{ii}]: The most prominent line is H$\alpha$+[\ion{N}{ii}]. It is seen in emission in the rim of both clouds. In the core spectra, the 
H$\alpha$ line is seen in absorption for both clouds. With our spectral resolution the H$\alpha$ line includes the [\ion{N}{ii}] lines at 6548~\AA\ and 6583~\AA, which were detected in emission in the diffuse galactic light by \citet{bd12}.   \\ 
$[\ion{S}{ii}]$: The second-most prominent line is [\ion{S}{ii}], including two lines at  6716~\AA\ and 6731~\AA, which are unresolved with our resolution.  In TPN the [\ion{S}{ii}] line is absent in the rim spectrum and negative in the core spectrum.  In DC~303 the [\ion{S}{ii}] line is positive in the rim spectrum and negative in the core spectrum.  \\
H$\gamma$ + G band: These two features are blended in our spectral resolution. In TPN these lines are not prominent, but the ratio spectrum shows a negative peak close to the G band. In DC~303 the lines can be seen in absorption in the rim spectrum, but are absent from the core spectrum. The ratio spectrum shows a wide negative peak ranging from $\sim$4250--4500~\AA\ that cannot be explained solely with H$\gamma$ and the G band. The Fraunhofer
G band is due to absorption by the CH molecule, with contribution from other molecular bands and metals.  \\
H$\beta$: This line is seen in absorption in the rim and core spectra of TPN and in absorption in the rim spectrum of DC~303, while it is absent from the core spectrum of DC~303.            \\
MgI + MgH: This band is seen in absorption in all spectra except the core spectrum of DC~303, from which it is absent.                  \\
Na D1+D2: The unresolved combination of these lines is in absorption in all spectra except the core spectrum of the TPN.         \\
TiO bands: These bands are in absorption in all spectra.       \\
\ion{Ca}{ii} triplet: The blend of the 8498 and 8542~\AA\ lines of this triplet is in absorption in all spectra. The third line at 8662~\AA\ is disturbed by a strong airglow band.           \\

A compilation of the spectral features is given in 
Table~\ref{tab:LineDetection}. 
In addition, the spectra include features that are artifical because they are only one channel wide. The emission-like feature at $\sim$6900~\AA\ is caused by the edge of the telluric O$_2$ absorption band.

\subsection{Ratio of brigth rim and dark core spectra}
Because the main noise contributor is the same in the spectra of the bright rim and the core, the noise between them is correlated. Therefore, the noise is expected to be reduced for the ratio of the bright rim and the core spectra.  Furthermore, for the TPN the rims are located symmetrically around the core, thus any imperfections in the correction of geometrical distortions (see Sect.~3) are largely cancelled out in the ratio.  Because the intensity of the core spectrum crosses zero, we derived the ratio in the following way: i) We fitted the spectra with a Chebyshev function, separately for wavelengths below and above the 4000~\AA\ break. ii) We subtracted the fitted function from the spectra and scaled the spectra so that the mean value of the rim and core spectra is unity. iii) Finally, we divided the rim and core spectra.  


\section{Modelling scattered light from globule and adjacent sky}

\subsection{Qualitative aspects}

To explain qualitatively the behaviour of scattered light at the bright rim and core of the TPN and DC~303 the following facts are to be realised.

i) The sky adjacent to the globules has a substantial dust column density. Accordingly, the background sky has continuum and line emission caused by scattering  of the interstellar radiation field, plus a signal from the {\em in situ}  gas emission. The extinction for the sky areas was determined using 2MASS \citep{skru06} {\it JHK} photometry and the NICER method of 
\citet{la01}: $A_V\mathrm{(sky)} \approx 1$~mag for both the TPN and DC~303.  The extinction and scattered light in front and behind of the globules have  to be considered separately. The foreground extinction layer can be assumed to be homogeneous over the slit length. Then, its contribution to the scattered light difference is cancelled out but influences the measured signal via extinction. The location of the  globules with respect to the extinction layer is not known.  Because most of the dust in this direction is located at the  distance of the Chamaeleon cloud complex \citep{k10}, we assumed that the globules are located in the middle of this layer,  with equal fore- and background extinctions of 
$A_V=0.5$~mag.

ii) At the position of the bright rim the optical depth through the globule, 
$\tau(\lambda)$, has a value of $\sim$2~mag. The conditions are therefore ideal for the scattering process.  At the very shortest wavelengths of our observations the intensity of the rims above sky background approaches zero.  This is mainly because the scattered light intensity  at the sky position is increasing towards shorter wavelenghts proportionally to  
$1-e^{-\tau(\lambda)}$, while the brigth rim signal is saturated and remains proportional to the ISL spectrum. A decreasing albedo of dust  particles may also contribute.

iii) At the position of the cores the optical depth through the TPN and DC~303 is much higher than unity, $\tau(\lambda)> 10$, at all wavelengths  covered by our spectra. The scattering process is therefore influenced by multiple scattering (and absorption) and the intensity of the scattered  light is consequently reduced. Furthermore, the light from behind the core is blocked. Thus, the surface brightness of the core relative to the sky becomes negative below a threshold wavelength. For the TPN this happens at 
$\lambda \sim 4300$~\AA, for DC~303 at $\lambda \sim 5900$~\AA.

iv)  The scattering geometry is different for the bright rims and the dark core. Interstellar grains are known to have a strongly forward-directed scattering function in the optical band, with the commonly used Henyey-Greenstein parameter \citep{hg41} in the range $g=0.7-0.8$ 
\citep{m70b,w90,d03}. This means that for the sky areas as well as for the bright rims the illumination is coming predominantly from behind the globule.  On the other hand, the increasing  optical thicknes  has been shown by 
\citet{m70b}  to modify the ``scattering function of the cloud'' from forward directed towards a more isotropic one. Thus, for the dark cores that are optically thick ($A_V \geq 10^\mathrm{m}$) the illumination is coming more uniformly from all over the sky or even from the backward direction.

\subsection{Nature of the H$\alpha$ emission from the globules: scattered 
            light}

If the  H$\alpha$ and other line emissions from the globule rims were caused by an ionised layer on the surface of the globule, one would expect to see a narrow intensity peak at the outer surface. On the other hand, if the emission is mainly scattered light, its distribution should be similar to the distribution of the adjacent continuum. We show in Fig.~\ref{fig:profiles} the surface brightness profiles of a band containing the H$\alpha$ and 
[\ion{N}{ii}] lines, together with the underlying continuum, and a band of the adjacent continuum.  Also shown is the ratio of these two bands accross the bright rims in each of the globules. Clearly, there is no hint of a thin ionised gas layer at the outer boundaries of the globules. The profiles over the bright rims are consistent with the idea that the H$\alpha$ and 
[\ion{N}{ii}] line emission in the bright rims is also scattered light.

\subsection{Modelling the continuum spectrum with ISL}

Because the observed spectra are differences between the globule and the nearby sky, the large foreground components of  the sky brightness, the zodiacal light, airglow, and tropospheric scattered light are eliminated. 
   
As explained in the appendix, we calculated for the solar neighbourhood  synthetic integrated starlight (ISL) spectra at different galactic latitudes and for locations at the Galactic plane ($z=0$) and at the height $z=50$~pc corresponding to the location of the two globules. 
 
The scattered light spectra of the globule rim and core are modelled as
\begin{equation}
I_{sca}^{cl}(\lambda) =  
I_{ISL}(\lambda)C_{5500}e^{-\tau_{V}^{eff}(cl)[a(\lambda)-a(5500 \AA)]} 
e^{-\tau_V(fg)a(\lambda)}, 
\end{equation}
where  $I_{ISL}(\lambda)$ is the ISL intensity, $a(\lambda)$ is the interstellar extinction \citep{c89} for $R_\mathrm{V}=3.1$ normalised to 1 at 5500~\AA, $\tau_V(fg)$ the foreground extinction and $\tau_{V}^{eff}(cl)$ the effective extinction of the cloud (rim or core) at 5500~\AA\ in optical depth units. $C_{5500}$ is a scaling factor to adjust the intensity level to the observed spectrum.

The scattered light intensity of the background dust layer at the 
sky position is modelled as 
\begin{equation}
I_{sca}^{sky,bg}(\lambda) = 
I_{sca}^{sky,bg}(5500 \AA)
\frac{I_{ISL}(\lambda)(1 - e^{-\tau_{V}(bg)a(\lambda)})}{I_{ISL}(5500 \AA)
(1 - e^{-\tau_{V}(bg)})} e^{-\tau_{V}(fg)a(\lambda)},  
\end{equation}

where $\tau_{V}(bg)$ is the extinction at 5500~\AA\ of the background dust layer in optical depth units. The observed continuum surface brightness globule-minus-sky is then given by \\ 
$ \Delta I_{sca}(\lambda)=I_{sca}^{cl}(\lambda)-I_{sca}^{sky,bg}(\lambda)$. \\
The sky background intensity at 5500~\AA, $I_{sca}^{sky,bg}$(5500~\AA), has to be estimated independently. We used two different methods:

(i) For an opaque core the scattered light in the ultraviolet, 
$\lambda \la 4000$~\AA, approaches zero. This is expected to be the case especially for DC~303, which has a core extinction of $A_V>50$~mag. The observed $\Delta I_{sca}$ at 4000~\AA\ is $\sim -8$~cgs. After correcting for the foreground extinction of 0.7~mag at 4000~\AA\ we obtain 
$I_{sca}^{sky,bg}$(4000~\AA)=15~cgs, and the same value for 
$I_{sca}^{sky,bg}$(5500~\AA).

(ii) We also used our surface photometry results for the high galactic latitude dark cloud L~1642 ($l=210\deg$, $b=-37\deg$) \citep{lms87, m11, m12}. In contrast to the TPN and DC~303, the surrounding sky of L~1642 is almost free of dust, $A_V \approx 0.15$~mag, 
$I_{sca}^{sky}(5500~\AA) \approx 3.4$~cgs. Plots of $I_{sca}(\lambda)$ vs.\ 
$A(\lambda)$,  corresponding to the $I_{sca}$ vs.\ $\tau_{100\mu\mathrm{m}}$ plots in \citet{lms87}, Fig.~7, show a linear increase of $I_{sca}(\lambda)$ for $A(\lambda) \le 1$, a flat maximum reached at 
$A(\lambda) \approx 1.5-2$~mag, and a decreasing surface brightness at  
$A(\lambda) \ga 2$~mag. We used these curves to derive for L~1642 the differences \\ 
$\Delta I_{sca}(\mathrm{L~1642})=I_{sca}(\mathrm{maximum})-I_{sca}(A_V=0.5$~mag$)$, \\ which correspond to the differential measurement ``bright rim minus sky'' for the TPN and DC~303. Comparing $\Delta I_{sca}(\mathrm{L~1642})$ values at the four wavelengths 3850, 4150, 4670, and  5470~\AA\ with the corresponding bright rim values for the TPN, we find an almost constant scaling factor of $1.5\pm 0.1$ which, when corrected for the foreground extinction $A_V = 0.5$~mag, corresponds to $1.9 \pm 0.13$. Using this scaling factor and the value of 7~cgs for $I_{sca}$(5500~\AA) at $A_V=0.5$~mag in L~1642, we obtain for the TPN background sky brightness the estimate  
$I_{sca}^{sky,bg}(5500\AA)=1.9 \times 7$~cgs=13~cgs.

\begin{figure}[htbp]
     \centering
     \begin{minipage}[t]{1.0\hsize}
       \centering
   \includegraphics[width=0.7\linewidth,angle=270]{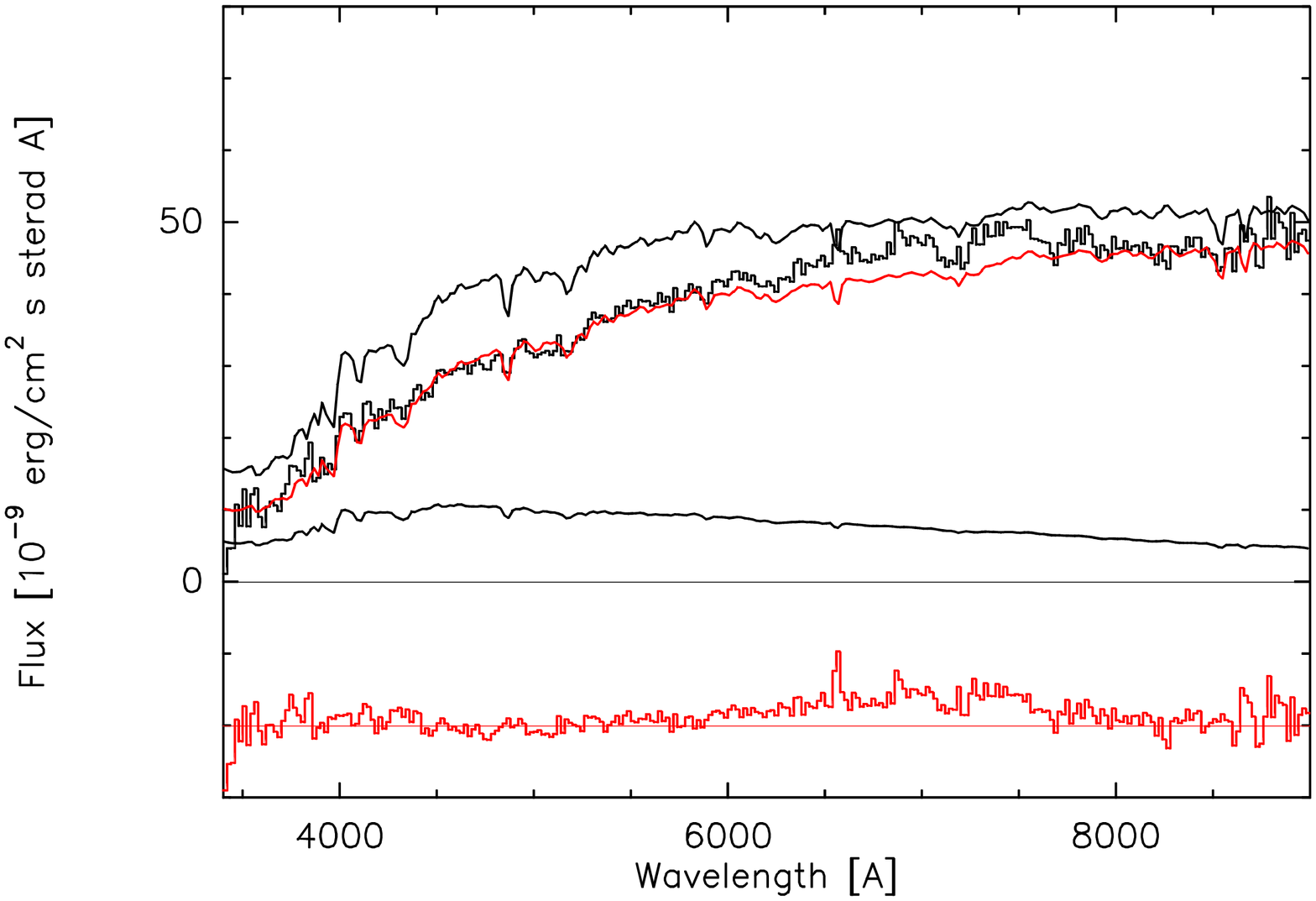}
     \end{minipage}%
     
     \begin{minipage}[t]{1.0\hsize}
        \centering
    \includegraphics[width=0.7\linewidth,angle=270]{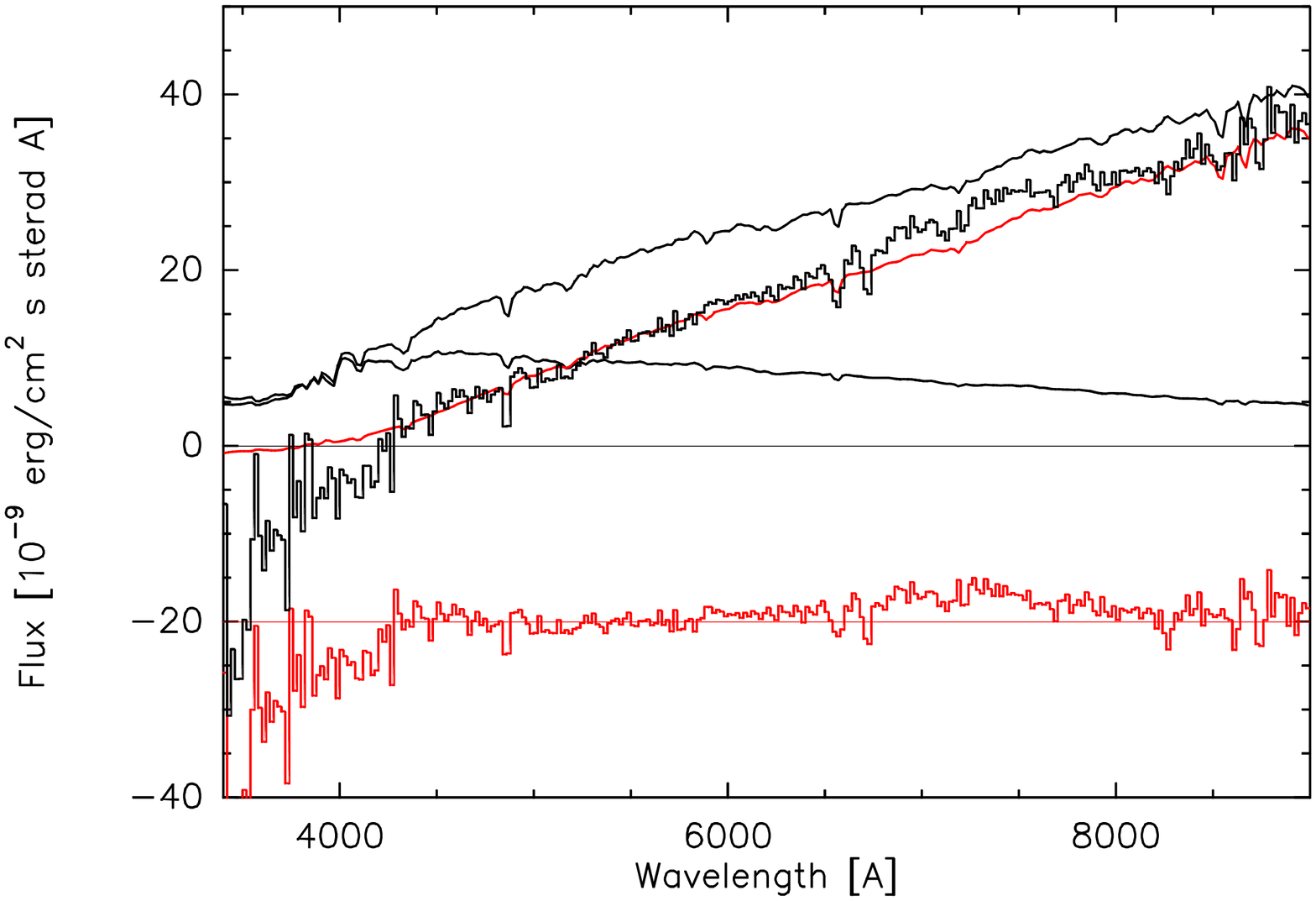}
     \end{minipage}  
    \caption{Fitting of the TPN spectra with ISL model spectrum. {\em The upper panel} is for the bright rim. The observed spectrum is shown as a histogram (thick black line). The upper and the lower continuous thin line in black show ISL model spectra for the bright rim and the sky positions, respectively. Their difference, the red continuous line, is the fit. The parameters used for the model spectrum were  
$A_{V}^{eff}(\mathrm{cloud})=1.086 \tau_{V}^{eff}(\mathrm{cloud})=0.33$~mag and $C_{5000}$=0.66. The residuals observation-minus-fit are shown as histogram (in red), shifted down by 20 units. {\em The lower panel} shows the core. Same as for the rim but the parameters used for the model fit were 
$A_{V}^{eff}(\mathrm{cloud})=1.086 \tau_{V}^{eff}(\mathrm{cloud})=1.4$~mag 
and $C_{5000}$=0.33.} 
    \label{fig:TPNfits}
    \end{figure}

\begin{figure} 
     \centering
     \begin{minipage}[t]{1.0\hsize}
       \centering
   \includegraphics[width=0.7\linewidth,angle=270]{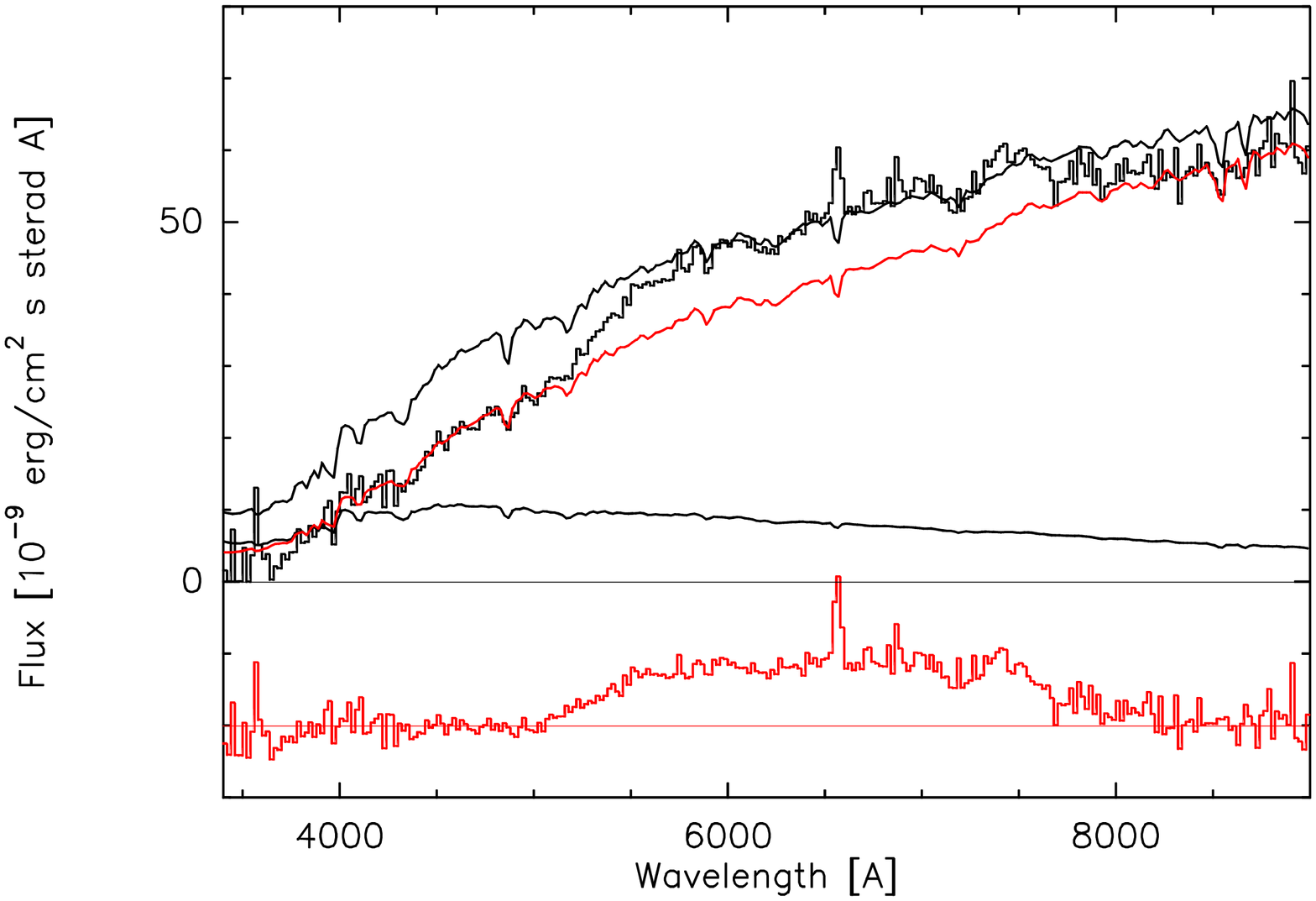}
     \end{minipage}%
     
     \begin{minipage}[t]{1.0\hsize}
        \centering
    \includegraphics[width=0.7\linewidth,angle=270]{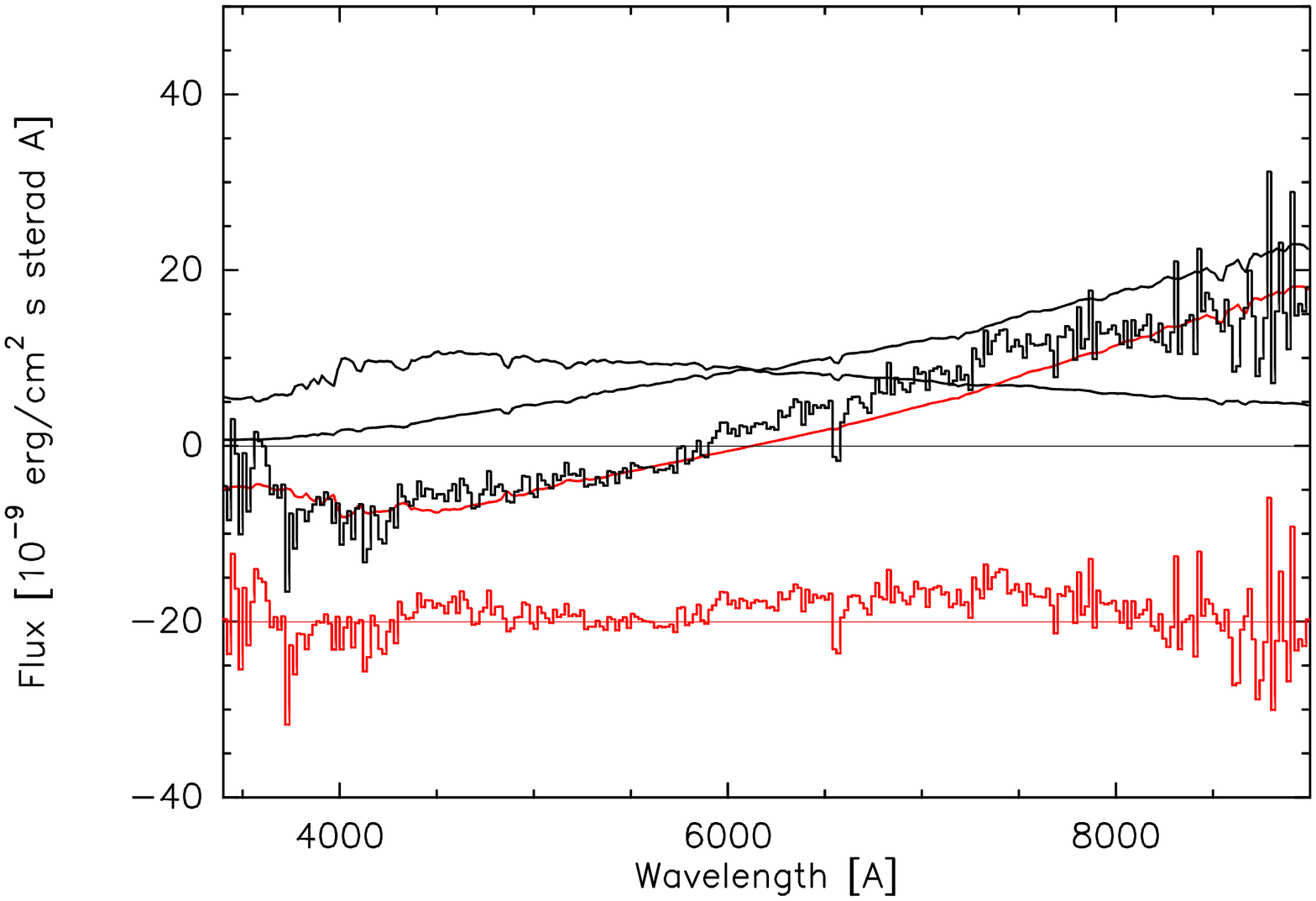}
     \end{minipage}
\caption{Fitting of the DC~303 spectra with ISL model spectrum. Same as for the TPN in Fig.~\ref{fig:TPNfits} but the fit parameters were 
$A_{V}^{eff}(\mathrm{cloud}) =1.086 \tau_{V}^{eff}(\mathrm{cloud})=1.05$~mag  and $C_{5000}$=0.60 for the rim (upper panel) and 
$A_{V}^{eff}(\mathrm{cloud})=1.086 \tau_{V}^{eff}(\mathrm{cloud})=2.7$~mag and $C_{5000}$=0.10 for the core (lower panel).} 
\label{fig:DC303fits}
\end{figure}

We show in Figs.~\ref{fig:TPNfits} and \ref{fig:DC303fits} the fitting of the observed spectra of the TPN and DC~303 with ISL spectra. For each figure the upper panel shows the fit for the bright rim and the lower panel for the dark core. The sky background spectrum was determined according to the recipe as described above and is fixed to the value $I_{sca}^{sky,bg} = 15$~cgs at 
5500~\AA\ for both globules.  The  ISL spectrum used for the bright rims is the mean of the ISL spectra at $|b|$=11.5 and 23.6~deg. For the dark cores the ISL spectrum is the mean over the sky. The values of $C_{5500}$ and 
$\tau_{V}^{eff}$(cl) in Eq.~(1) are adjusted to optimise the fits to the observed spectra and are given in the figure legends. Residuals for the observed-minus-model spectrum are shown in the lower parts of the figures. For the DC~303 rim and more weakly for the TPN rim an excess emission above the scattered light model spectrum is found between $\sim$5000 and 8000~\AA. 
Its properties and interpretation are discussed in Sect.~6.4.

\subsection{Scattered and {\em in situ} emission lines}

Scattered emission line radiation from all over the sky contributes to the rim and core spectra of the globules as well as to the spectrum of the sky position in a similar  way as the ISL continuum light. In addition, there is {\em in situ} line emission  from ionised gas in front and behind the globules. The emission lines to be considered here are 
H$\alpha$+[\ion{N}{ii}]$\lambda$6548,6583, H$\beta$, and 
[\ion{S}{ii}]$\lambda$6716,6731.  

These lines, after subtracting the integrated starlight model spectra,
are found to be positive (in emission) or zero in the rim and negative (apparently in absorption) in the core spectra, because emission lines from the sky position dominate in the latter case. We denote with $I_{sca}^l$ the {\em scattered} line intensity integrated over the emission line profile and with $I_{in situ}^l$ the intensity of the {\em in situ} line emission from behind the globule.  The {\em in situ} as well as scattered line emission in front of the globule is assumed to be constant over the 6 or 8 arcmin slit length and thus to cancel out in the differential measurements.  It is meaningful to combine the scattered line component of the background sky behind the globule, $I_{sca}^l$(sky), with the rim or core  scattered line component because it derives its illumination from the same source, the all-sky line radiation dominated by the bright HII regions in the Galactic plane. On the other hand, the {\em in situ} background line emission is produced in the local warm ionised medium (WIM), which has different physical conditions.  The two components have different emission line ratios \citep{m06}.  

We assumed that the optical depth through the bright rim at the wavelength of each line is $\tau$(rim)=2 while the core is assumed to be opaque,  
$\tau$(core)$>10$, transmitting practically no background light. The optical depth of the foreground dust layer is denoted with $\tau$(fg) and corresponds to the dust extinction of $A_V$=0.5~mag as adopted in Sect.~5.1.

The scattered component of the line intensity globule-minus-sky can be expressed as 
\begin{eqnarray*}
\Delta I_{sca}^l(rim)&=&[I_{sca}^l(rim) + I_{sca}^l(sky) 
       e^{-\tau(rim)}-I_{sca}^l(sky)]e^{-\tau(fg)}  \\
\Delta I_{sca}^l(core)&=&(I_{sca}^l(core)-I_{sca}^l(sky))e^{-\tau(fg)}.  
\end{eqnarray*}
We furthermore assumed that the equivalent width, EW, for a given line, e.g.\  H$\alpha$, is the same for each scattered line component, i.e.\ rim, core and sky background and, that it is also the same for the TPN and DC~303,

\begin{eqnarray*}
I_{sca}^l(rim)&=&EW \times I_{sca}^c(rim)     \\ 
I_{sca}^l(core)&=&EW \times I_{sca}^c(core)   \\ 
I_{sca}^l(sky)&=&EW \times I_{sca}^c(sky), 
\end{eqnarray*}
where the continuum intensity $I_{sca}^c$ is at the wavelength of the emission line. Thus we can write
\begin{eqnarray*}
\Delta I_{sca}^l(rim)&=&EW \times \Delta I_{sca}^c(rim) \\
\Delta I_{sca}^l(core)&=&EW \times \Delta I_{sca}^c(core).    
\end{eqnarray*}
The observed line intensity (scattered + {\em in situ}) difference is given by  
\begin{eqnarray*}
&\Delta I^l(rim )&=EW\times \Delta I_{sca}^c(rim) + I_{in situ}^l 
        (e^{-\tau(rim)} - 1)e^{-\tau(fg)} \\
&\Delta I^l(core)&=EW\times \Delta I_{sca}^c(core) - I_{in situ}^l 
        e^{-\tau(fg)}. 
\end{eqnarray*}
In these equations the diffential continuum intensities (rim minus sky and core minus sky), 
$ \Delta I_{sca}^c(rim)$  and  $\Delta I_{sca}^c(core)$, 
represent the scattered light values fitted to the observations (red curves in Figs.~\ref{fig:TPNfits} and \ref{fig:DC303fits}) and they were previously corrected for the foreground extinction $\tau$(fg).

Given the observed line intensities $\Delta I^l$(rim) and $\Delta I^l$(core)  for the TPN and DC~303, we solved the group of four equations for the three unknowns, $EW$, $I_{in situ}^l$(TPN), and $I_{in situ}^l$(DC~303). The results are given in columns 6 and 7 of Table~\ref{tab:LineIntensities}. The three scattered line components $I_{sca}^l(rim)e^{-\tau(fg)}$,  
$I_{sca}^l(core)e^{-\tau(fg)}$, and $I_{sca}^l(sky)e^{-\tau(fg)}$ were calculated using the $EW$ and the corresponding scattered continuum components. They are listed in columns 8--10 of 
Table~\ref{tab:LineIntensities}.

\section{Discussion}

\subsection{Milky Way ISL spectrum in scattered light}

The SED of the bright rim of a globule is expected to be only weakly infuenced by interstellar reddening in the cloud, i.e.\ that it is almost a direct copy of the SED of the impinging ISRF in spite of the substantial optical thickness. This is true as long as the albedo and scattering function asymmetry parameter $g$ are constant over the wavelength range in question. This appears to be a good approximation over the range considered here, $\lambda=3500-9000\AA$ \citep{m70b,lms87,d03}.  Thus, we expect that the bright rim spectrum is a good tracer of the ISL spectrum not only for the Fraunhofer lines and spectral discontinuities, but also regarding the shape of the SED. 

The SED of the TPN rim comes close to this expectation: the best-fit ISL model spectrum (with $z=50$~pc) is obtained with a reddening correction corresponding to $A_{V}^{eff}$(cl)=$1.086 \tau_{V}^{eff}$(cl)=0.33~mag or  
$E(B-V)=0.11$~mag. This reddening can easily be explained with the slightly decreasing albedo, from 0.52 at 5500~\AA\ to 0.45 at 3450~\AA\ \citep{lms87}.

The SED for DC~303 rim is clearly redder than for the TPN, corresponding to  
$A_{V}^{eff}$(cl)=1.05~mag or $E(B-V)=0.34$~mag. Since the TPN and DC~303 are exposed to about the same radiation field, the different rim reddening is probably caused by different dust properties. A difference in dust composition is also suggested  by the presence of a stronger ERE component in DC~303 as compared to the marginal  detection in the TPN. Another reason could be the different density distributions of the two globules \citep{k07}, DC~303 being more centrally peaked than the TPN. A detailed radiative transfer modelling would be required to study this effect, but this is beyond the scope of this paper.

The scattered light intensities for the TPN and DC~303 rims are very similar, as can be seen from Figs.~\ref{fig:TPNfits} and \ref{fig:DC303fits} and the scaling factors, $C_{5500}$=0.66 and 0.60 for the TPN and DC~303, respectively. 

The bright rim spectra show the stronger Fraunhofer lines with very similar strengths as observed in the ISL model spectrum. The 4000~\AA\ discontinuity is also pronounced. We regard that this, as well  as the agreement between the observed and modelled SEDs, supports the choice of parameters used in our population synthesis model for the ISL.  One possible deviation may be indicated for the TiO band at  $\sim$7100~\AA. The observed band strength appears to be somewhat stronger than predicted, but the presence of ERE covering this wavelength range may be an explanation as well. This band is sensitive to metal abundance; thus, adding more metal-rich late-type giants might improve the fit. 
 
For the dark cores of the TPN and DC~303, where $A_V \geq 10^m$, the scattered light SED is strongly influenced by multiple scattering and absorption. Between the subsequent scattering events the light experiences wavelength-dependent extinction and the scattered light becomes strongly reddened. The reddening law as determined for point sources is not valid  for diffuse radiation penetrating through a scattering medium  \citep{m76}. For our modelling purpose, however, we formally used the point source reddening curve to fit the ISL model SED to the observed spectra. As shown in 
Figs.~\ref{fig:TPNfits} and \ref{fig:DC303fits}, good fits can be achieved this way. The formal effective extinction values adopted were 
$A_{V}^{eff}$(cl)=$1.086 \tau_{V}^{eff}$(cl)=1.4~mag and 2.7~mag, and the scaling factors $C_{5500}$=0.33 and 0.10 for the TPN and DC~303, respectively. 

Because the surface brightness at the sky position becomes a substantial fraction of, or even exeeds, the core surface brightness, the scattered Fraunhofer lines become weak in the core-minus-sky spectra. On the other hand, because the cores are opaque to the background light, their spectra show the emission lines of the {\em in situ} gas at sky positions as apparent absorption lines.   

\subsection{Emission lines: comparison with other studies}

Equivalent widths of the scattered emission lines according to \citet{bd12} are presented in the lower part of Table~\ref{tab:LineIntensities}. The agreement with our value for H$\alpha$+[\ion{N}{ii}]$\lambda 6583$,$\lambda 6548$ is surprisingly good considering the different approaches of the two studies. The [\ion{S}{ii}]$\lambda 6716$,$\lambda 6731$ equivalent width of \citet{bd12},  on the other hand, is significantly larger than our value. Two reasons can play a role: (1) The illumination for the scattered line radiation is coming from different areas of sky, predominantly from behind the globules in our case and more uniformly all over the sky in \citet{bd12}; (2) As stated by  \citet{bd12} there may be a contribution by {\em in situ}  gas emission correlated with the 100 $\mu$m fluctuations in their values, which increases the apparent scattered light signal. Supporting this, a correlation of 
H$\alpha$ with FIR dust emission has been claimed by \citet{l99} to be evidence of a first detection of dust emission by WIM. \footnote{However, these authors omitted the scattered H$\alpha$ radiation, which may have been the reason for the detected correlation.} 

We give in Table~\ref{tab:LineIntensities} also the 
H$\alpha$+[\ion{N}{ii}]$\lambda 6583$,$\lambda 6548$  equivalent width as measured by \citet{w10} in the high-latitude dark nebula L~1780. Their 60~\AA\ wide filter centred at the H$\alpha$ line encompassed also the two [\ion{N}{ii}] lines. The [\ion{S}{ii}]$\lambda 6716$,$\lambda 6731$ equivalent width was derived using line ratio measurements discussed in Sect.~6.3 below. The  
H$\alpha$+[\ion{N}{ii}]$\lambda 6583$,$\lambda 6548$ equivalent width is somewhat smaller than found by us from the TPN and DC~303 and by \citet{bd12} from scattering by diffuse dust. Part of this difference may be caused  by the difference in calibration of our spectroscopic and their narrow band photometric data. Another reason could be the background {\em in situ}  emission, which was not explicitly handled in the analysis of \citet{w10}. The equivalent width of 4.5~\AA\ for [\ion{S}{ii}]$\lambda 6716$,$\lambda 6731$ is compatible with our value of $3.7\pm1.1$ and significantly lower than the value of \citet{bd12}. 

The equivalent width determined for the scattered  
H$\alpha$+[\ion{N}{ii}]$\lambda 6583$,$\lambda 6548$ emission in the two globules is 20.2~\AA. The mean continuum intensity of the Milky Way from Pioneer 10 red photometry (with $m_V<6.5$~mag stars included) is 
167~S10$_{\sun}$ \citep{t81}, corresponding to 173~cgs at 6560~\AA. Thus the 
H$\alpha$+[\ion{N}{ii}] $EW$ of 20.2~\AA\ corresponds to 3490~cgs\AA\ or 14.5~R. Assuming that 26\% of this signal is due to 
[\ion{N}{ii}]$\lambda 6583$,$\lambda 6548$  as found by \citet{m06} for HII regions, we obtain for the mean H$\alpha$ intensity the value of 11.5~R, which agrees reasonably well with the mean intensity of 8.0~R as found from the 
H$\alpha$ map of \citet{f03}. 

In the SHASSA survey \citep{g01} the H$\alpha$ sky intensities  next to the two globules are 6.4~R (1540~cgs\AA) for the TPN and 7.2~R (1730~cgs\AA) for DC~303. These values refer to the whole line of sight.  Our estimates,  referring to H$\alpha$ {\em in situ} emission from behind the globules, are 
56\% and 32\% of these values. This is in accordance with our assumption that the globules are embedded half way within the dust and gas layer towards the Chamaeleon complex.

\subsection{Emission line ratios}

We list in Table~\ref{tab:LineRatios} the observed line ratios for the scattered and {\em in situ} line components. The [\ion{N}{ii}]/H$\beta$ ratio for the {\em in situ} case was calculated from  
(H$\alpha$+[\ion{N}{ii}])/H$\beta$ by assuming that H$\alpha$/H$\beta$=2.87 corresponding to case B for $T$=10000~K and no extinction \citep{of06}. Likewise, [\ion{N}{ii}]/H$\alpha$ was calculated from 
[\ion{N}{ii}]/H$\beta$ using the same H$\alpha$/H$\beta$ ratio.

For comparison we give in column 6 the line ratios according to \citet{bd12} for the scattered light from diffuse dust. \citet{m06} have made extensive observations  of the line ratios [\ion{N}{ii}]$\lambda 6583$/H$\alpha$ and [\ion{S}{ii}]]$\lambda 6716$/H$\alpha$, both for bright HII regions and for high-latitude WIM gas. To compare their results with ours, we included both line components of [\ion{N}{ii}] and [\ion{S}{ii}] using the ratios  
[\ion{N}{ii}]$\lambda 6583$/$\lambda 6548$=2.97 and 
[\ion{S}{ii}]$\lambda 6716$/$\lambda 6731$=1.45.

\citet{w10} have reported measurements with the Wisconsin H-alpha Mapper (WHAM) of the high galactic latitude dark nebula L~1780. The line ratios of the excess surface brightness averaged over the one-degree WHAM beam were  
[\ion{S}{ii}]$\lambda 6731$/H$\alpha$=0.19 and 
[\ion{S}{ii}]$\lambda 6731$/H$\alpha$=0.16. Using again the ratios  
[\ion{N}{ii}]$\lambda 6583$/$\lambda 6548$=2.97 and 
[\ion{S}{ii}]$\lambda 6716$/$\lambda 6731$=1.45 we obtain for 
[\ion{S}{ii}]$\lambda 6716$,$\lambda 6731$/(H$\alpha$,[\ion{N}{ii}]$\lambda 6583$+$\lambda 6548$) the value of 0.31 (column 9 of Table~\ref{tab:LineRatios}). This is the same value as found by us for scattered light in the TPN and DC~303.

The line ratios [\ion{N}{ii}]/H$\alpha$ and [\ion{S}{ii}]/H$\alpha$ differ systematically for bright HII regions and the diffuse high-latitude WIM gas \citep{m06}. Our observed ratio 
[\ion{S}{ii}]$\lambda 6716$,$\lambda 6731$/(H$\alpha$+[\ion{N}{ii}]$\lambda 6583$,$\lambda 6548$), is 0.31 for the scattered and 0.51 for the {\em in situ} component. These values  fall in between the \citet{m06} values for HII regions (0.14) and WIM gas ($\sim 0.7$, with substantial variations). While  the scattered line radiation is dominated by illumination from HII regions, a substantial part of it is likely coming from the faint but widely distributed WIM, which causes the enhanced line ratio above the HII regions value of 0.14. The value of \citet{bd12}, supposed to be dominated by scattered radiation  from HII regions, is substantially higher and almost the same as  our value for the {\em in situ} emission. As stated above, this may indicate that part of the [\ion{S}{ii}] line emission detected by \citet{bd12} is not scattered, but {\em in situ} emission.

Our (H$\alpha$+[\ion{N}{ii}])/H$\beta$ ratio is higher for the scattered than for the {\em in situ} line emission, likely because of the interstellar reddening of the light from HII regions. The \citet{bd12} value is consistent with our scattered light ratio.  Our [\ion{N}{ii}]/H$\beta$ and [\ion{N}{ii}]/H$\alpha$ ratios have substantial scatter.  They are consistent with the HII regions value of \citet{m06} but not with the WIM.  The \citet{bd12} value, on the other hand, is consistent with the WIM but not with the HII regions origin. 

We conclude that the emission line spectra of the bright rims and dark cores of the TPN and DC~303 can be modelled as a combination of  external illumination, leading to scattered light from the globules, plus {\em in situ}  emission by gas  behind the globules.  In the case of the H$\alpha$ line, \citet{mjl07} and \citet{ljm10} have previously interpreted several observed dark cloud maps in terms of combined scattered plus {\em in situ} emission. \citet{w10} have used the low [\ion{S}{ii}]/H$\alpha$ ratio observed in L~1780 as supporting evidence for the scattered light origin of most of the H$\alpha$ luminosity of this cloud.

\begin{table*}
 \centering
\caption{Scattered and {\em in situ} line emission derived from the TPN and DC~303 spectra. In columns 8--10  the numbers in parentheses are the corresponding scattered continuum intensities in cgs units.The [\ion{N}{ii}] line strengths include $\lambda 6583$ and $\lambda 6548$ lines and [\ion{S}{ii}] includes $\lambda 6716$ and $\lambda 6731$ lines. In the lower part of the table the line equivalent widths are given as observed by \citet{bd12} for diffuse dust and by \citet{w10} for the dark nebula L~1780.}
  \begin{tabular}{@{}lccccccccccc@{}}
  \hline 
  Line & $\Delta I^l$(rim) & $\Delta I^l$(core) & $\Delta I_{sca}^c$(rim) & 
  $\Delta I_{sca}^c$(core) & $EW$ & $I_{in situ}^l$
&  \multicolumn{3}{c}{$I_{sca}^le^{-\tau(fg)}$}\\
 &  & & & & & & (rim) & (core) & (sky)\\
                     &  \multicolumn{2}{c}{cgs \AA\ \tablefootmark{a}} & \multicolumn{2}{c}{\small cgs} & \AA & \multicolumn{4}{c}{\small cgs\AA}  \\
(1) & (2) & (3) & (4) & (5) & (6) & (7) & (8) & (9) & (10) \\
 \hline
TPN      &        &       &         &        &       &      &    \\        
 H$\alpha$+[\ion{N}{ii}] & $234\pm27$ & $-228\pm32$ &  42     & 20.0   &$20.2\pm0.85$ & $967\pm54$ & 1060(50) & 424(20) & 170(8)\\
  $[\ion{S}{ii}]$     & $0\pm27$  & $-241\pm32$ &  42.5   & 22.5   & $6.29\pm0.83$  & $502\pm55$ & 275(50) & 121(22) &  44(8)\\
 H$\beta$       & $0\pm42$   & $-148\pm32$ &  32     &  7.0   & $3.74\pm1.10$ & $274\pm55$ & 130(42) &  53(17) &  28(9)\\
 H$\alpha$  \tablefootmark{b} &  &       &         &        &       & $786\pm158$ & & & & \\
\hline
DC~303      &        &       &        &      &      &      &     \\        
 H$\alpha$+[\ion{N}{ii}] & $559\pm29$ & $-309\pm28$ &  42.5   & 4.5  & $20.2\pm0.85$ & $548\pm42$ & 1081(51) & 233(11) & 174(8) \\
  $[\ion{S}{ii}]$  & $96\pm29$ & $-126\pm28$ &  43     & 5.0  & $6.29\pm0.83$  & $252\pm41$ &  286(52) &  63(11.5) & 44(8) \\
 H$\beta$       & $  0\pm28$ & $- 95\pm28$ &  25     & -5.0   & $3.74\pm1.10$ &$155\pm41$ &  108(35) &  31(10) &  28(9) \\
 H$\alpha$$^b$ &         &       &         &        &       &    $445\pm118$  &  &         &         & \\
\hline
\multicolumn{2}{l}{Brandt \& Draine (2012)} &       &         &        &       &      &     &         &         & \\
 H$\alpha$+[\ion{N}{ii}]  &       &       &         &        &  $21.5\pm0.9$ &      &     &         &         & \\
  $[\ion{S}{ii}]$       &       &       &         &        & $10.0\pm0.9$  &      &     &         &         & \\
 H$\beta$         &       &       &         &        & $4.8\pm0.7$   &      &     &         &         & \\
\hline
\multicolumn{2}{l}{Witt et~al.\ (2010)}      &       &         &        &       &      &     &         &         & \\ 
 H$\alpha$+[\ion{N}{ii}]  &       &       &         &        & 14.4  &      &     &         &         & \\
  $[\ion{S}{ii}]$       &       &       &         &        & 4.5  &      &     &         &         & \\
\hline
\end{tabular}
\tablefoot{ 
\tablefoottext{a}{cgs=\cgs }     
\hspace{1cm}
\tablefoottext{b}{assuming  H$\alpha$/H$\beta$=2.87}   }
\label{tab:LineIntensities}
\end{table*}

 \begin{table*}
 \centering
\caption{Scattered and {\em in situ} emission line ratios. The  
[\ion{N}{ii}]/H$\beta$ ratios for the {\em in situ} case were calculated by assuming a ratio  H$\alpha$/H$\beta$=2.87 corresponding to case B for 
$T=10000$~K and no extinction \citep{of06}. The [\ion{N}{ii}] line strengths include $\lambda 6583$ and $\lambda 6548$ lines and  [\ion{S}{ii}] includes  
$\lambda 6716$ and $\lambda 6731$ lines. In \citet{m06} and \citet{w10} the values are for single lines $\lambda 6583$ and $\lambda 6716$ and were modified here corresponding to the line ratios.}
  \begin{tabular}{@{}l|c|ccc|cccccc|@{}}
  \hline
Line ratio & {Scattered} & \multicolumn{3}{c}{In situ} & Brandt \& & \multicolumn{2}{c}{Madsen et al.\ (2006)} & Witt et al.\\
           & TPN \& DC~303 & TPN & DC~303 & Mean & Draine (2012) & HII reg & WIM & (2010) \\  
 (1) & (2) & (3) & (4) & (5) & (6) & (7) & (8) & (9)\\
\hline
  [\ion{S}{ii}]/(H$\alpha$+[\ion{N}{ii}])   & $0.31\pm0.05$ & 0.52  & 0.50 & 0.51$^{+0.07}_{-0.04}$ & 0.46$^{+0.06}_{-0.03}$ & 0.14 & 0.7 & 0.31 \\ 
 (H$\alpha$+[\ion{N}{ii}])/ H$\beta$  & 5.4$^{+2.6}_{-1.4}$   & 3.5  & 3.5 
 & 3.5$^{+0.8}_{-0.5}$  & 4.5$^{+1.0}_{-0.8}$  &  & & \\
 $[\ion{N}{ii}]$/ H$\beta$ \tablefootmark{a} & & 0.7 & 0.7 & 
 0.7$^{+0.8}_{-0.5}$ & 1.9$^{+0.5}_{-0.4}$ & & & \\
$[\ion{N}{ii}]$/ H$\alpha$ \tablefootmark{a} & & 0.2 & 0.2 & 
0.2$^{+0.3}_{-0.2}$ & 0.72$^{+0.09}_{-0.08}$  & 0.36 & 0.7 & \\
\hline
\end{tabular}
\tablefoot{ 
\tablefoottext{a}{assuming H$\alpha$/H$\beta$=2.87} }
\label{tab:LineRatios}
\end{table*}

\subsection{The excess emission at 5200--8000~\AA}

\begin{figure}
\resizebox{\hsize}{!}{\includegraphics[angle=270]{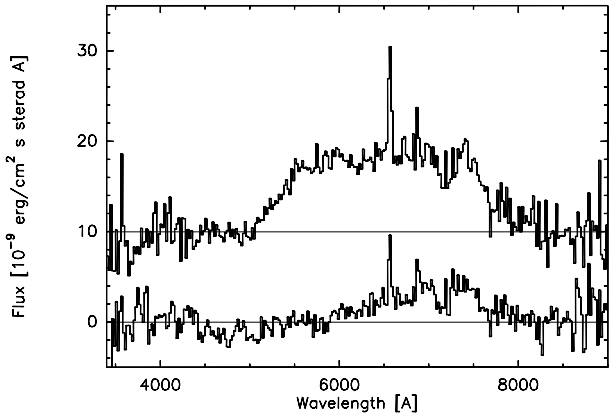}}
\caption{Excess emission from the bright rim of the TPN (lower spectrum) and DC~303 (upper spectrum). The spectrum for DC~303 was shifted upwards by 10~cgs. The H$\alpha$+[\ion{N}{ii}] line emission in both spectra and 
[\ion{S}{ii}] line emission in DC~303 are unrelated to the ERE.}
\label{fig:ERE}
\end{figure}

The residuals of the observed globule spectra fitted with the synthetic  ISL spectrum are shown in Fig.~\ref{fig:ERE}. For the DC~303 rim there is an excess emission of $\sim7$~cgs over the wavelength range 5200--8000~\AA\, while the TPN rim shows  weaker excess emission of $\sim3$~cgs over 6200--8000~\AA. Unrelated to this, strong H$\alpha$+[NII] line emission is seen in both spectra and weak [SII] line emission in DC~303, in addition. Even though the observed excess in the TPN might be  suspected to be caused by observational effects or imperfect ISL SED modelling, it is not likely that at the same time DC~303 would show a much stronger excess. If caused by problems in the ISL modelling, both globules should show very similar excesses. We first summarise its observed properties and then discuss some alternative interpretations.

The basic parameters characterising the emission feature are the central wavelength $\lambda_c$=6548~\AA, width $\Delta \lambda=2800$~\AA, band intensity 
$I_{\mathrm{emission}}=1.9 \times 10^{-5}$~\cgss 
for DC~303 and $\lambda_c=6950$~\AA, $\Delta \lambda=1800$~\AA, and   
$I_{\mathrm{emission}}=4.7 \times 10^{-6}$~\cgss 
for the TPN. Compared to the scattered light within the emission band wavelength  range,  $I_{\mathrm{emission}}/I_{\mathrm{sca}} \approx 0.15$ for DC~303 and 0.047 for DC~303 and the TPN, respectively. Emission band intensity and central wavelength were determined as the zeroth and first moments over the band; the width is an estimate made by eye. The emission band intensity values are the observed ones. If corrected for $A_V=0.5$~mag foreground extinction, they would increase to 2.7 and 0.67$\times 10^{-5}$~\cgss  
for DC~303 and the TPN, respectively. These values are similar to the estimate for the ERE band intensity in L~1780,  
$I_{\mathrm{ERE}} \approx 2\times 10^{-5}$~\cgss, 
as shown in Fig.~1 of \citet{sw02} and, if interpreted in the same way as these authors, will lead to a similarly high photon conversion efficiency of 
$\eta \approx 10\%$. Our emission band to scattered light ratio for DC~303 is lower by factor of $\sim4$ than the estimate of \citet{sw02} for L~1780. Half of this difference is due  to the stronger scattered light caused by the 
$\sim2$ times stronger illumination from ISL in the direction $|b|=15$deg of DC~303 and the TPN as compared to ISL in the direction $|b|=37$deg of L~1780.

The emission band intensities are, similarly as for the scattered light continuum, differences between globule rim and diffuse emission from the surrounding sky. Our observations do not allow us to put limits to the emission band intensity at the sky positions. In their analysis of the scattered light spectrum  of diffuse interstellar dust, \citet{bd12} have found no evidence for ERE emission.  

Strong ERE has been detected with good statistical and systematic accuracy in high surface brightness objects, such as reflection nebulae. Among the  more difficult ERE searches in faint surface brightness objects the present result has some advantages. It is based on subtracting a scattered light spectrum, which is more realistic than in the previous works. These were based either  on a general template galaxy spectrum as proxy for the local ISRF spectrum 
\citep{s98}, or on a few broad- or intermediate-band photometric  points only \citep{g98, w08}. The present search for ERE in the two globules can complement these previous determinations because of its wide spectral coverage and sufficient resolution. Especially, we suggest that our result is  observationally preferable to the low-resolution spectrophotometry of   L~1780 by \citet{m79}, which has been interpreted by \citet{cl87} to show ERE  emission and has later been used as the only representative of an 
ERE-bearing dark/translucent cloud so far \citep{sw02,gd10}.

It is {\em phenomenologically} suggestive to interpret the excess emission at 5200--8000~\AA\ as extended red emission rising from the photoluminescence process. However, the interstellar environment and the specific physical circumstances under which these globules were observed appear to speak against this interpretation:\\ 
(i) Observations of reflection nebulae with different types of illuminating stars have indicated that far-UV photons at wavelengths shortward of 1180~\AA\ are needed for ERE exitation \citep{darbon,witt1985,witt2006};\\ 
(ii) these photons can penetrate only a thin surface layer of the globules;\\ 
(iii) therefore, a uniform ERE band surface brightness profile over the globule's face is predicted rather than a bright rim - dark core structure as observed for DC~303 and the TPN \citep{w90};\\ 
(iv) because our sky positions with $A_V \sim 1$~mag are also optically thick at far--UV wavelengths, practically equal amounts of ERE emission are expected at sky and globule sight-lines, causing the {\it differential} measurement to deliver a null result independently of the actual strength of the ERE emission.      

However, it is possible to construct somewhat {\em ad hoc} a model for the dust distribution and UV illumination in the globule surroundings that would make the ERE interpretation of the excess emission in DC~303 look more plausible: \\
(i) Assuming that DC~303 is located behind most of the general $A_V \sim 1$~mag extinction layer as observed in the sky position. In this situation our differential measurement technique would not nullify the ERE signal because an equally high foreground ERE intensity is present both at the sky and the globule line of sight.\\ 
(ii) DC~303 would then be exposed freely to the far-UV radiation field escaping from the OB stars in the Galactic disk. This radiation is impinging on DC~303 from the north (see Fig.~1) and would be similar in strength to the ISRF that excites the ERE detected in the diffuse ISM \citep{g98}. \\
(iii) The missing ERE emission towards the dark core of DC~303 can be understood since the far-UV radiation field is asymmetric, coming mostly from the side (north) and to lesser extent from the observer's direction. In this case the dark core would receive less far-UV illumination than the rim and less than assumed in the globule model of \citet{w90}.

Two observational effects expected in this {\it ad hoc} model can be noticed: (i) the Galactic plane (north) side of DC~303, exposed to stronger far-UV radiation, is seen to be brighter than the south side in blue and red images (cf Fig.~\ref{fig:DSS-images}); (ii) if the TPN and DC~303 were both embedded in the surrounding $A_V \sim 1$~mag extinction layer the SEDs of their bright rim diffuse emission should be the same for both and equal to the SED of ISL. On the other hand, if DC~303 is behind the $A_V \sim 1$~mag extinction layer and the TPN in front or inside it, DC~303 should be more strongly reddened than the TPN. This is in fact the case: the colour of the bright rim of DC~303 is redder than that for the TPN, corresponding to an effective extinction of   $A_V \sim 1$~mag (see Figs.~\ref{fig:TPNfits} and \ref{fig:DC303fits}, upper panels). We also estimated the extinction in the bright rim areas of the TPN and DC~303 using 2MASS {\it JHK} photometry of background stars in the two globule areas. The areas were defined as having a red surface brightness on DSS red plates (see Fig.~\ref{fig:DSS-images}) of $>0.7 \times$ the maximum. The resulting R-band extinctions, determined from 
$\sim$24 stars in each globule, were $2.4 \pm 0.3$ and $3.0 \pm 0.3$ for the TPN and DC~303, respectively. These values are compatible with the model where the bright rim corresponds to an optical depth through the globule of 
$A_R \sim 2$~mag, and the foreground extinctions are  
$A_R \sim 0.4$~mag for the TPN (embedded in the $A_R \sim 0.8$~mag layer) and  $A_R \sim 1$~mag for DC~303 (behind the $A_R \sim 0.8$~mag layer).

Another alternative to explain the excess emission at 5200--8000~\AA\ is in terms of an increased albedo over just this wavelength range and more strongly in DC~303 than in the TPN. Although such an albedo behaviour has not been observed before, it is also true that there have been very few albedo determinations in dense clouds covering this wavelength range with sufficient spectral resolution. On the other hand, the observations of reflection nebulae with late A to M-type illuminating stars should have revealed, if present, such an albedo enhancement \citep{darbon}. For B and early A-type illuminating stars the ERE emission in reflection nebulae dominates this wavelength region and would hide such an albedo enhancement.

\section{Summary and conclusions}

We have obtained long-slit spectra 3500--9000~\AA\ of the faint 
surface brightness of two nearby ($d\sim200$~pc) globules associated
with the Chamaeleon dark cloud complex. The much stronger airglow emission 
spectrum and the atmospheric absorption lines were succesfully elimated 
over most of the wavelength range. Spectra were presented separately for
the bright rim and dark core regions of the globules. Although of relatively 
low resolution ($R\sim 100-200$), the spectra allowed us to derive 
a number of results:

The spectra are mainly caused by scattered light from the ISRF in the local Milky Way, consisting of a continuous/absorption line component of the integrated light of stars. The SED of the TPN bright rim, corresponding to an optical depth of $\tau \approx 2$,  is almost identical to the synthetic ISRF spectrum. Also, in all globule spectra the stronger Fraunhofer lines of the integrated starlight spectrum were detected, with the limitation imposed by the relatively low spectral resolution of our observations.  

In addition, emission lines of H$\alpha$+[\ion{N}{ii}], H$\beta$, and 
[\ion{S}{ii}] were detected and were interpreted in terms of scattered light from bright HII regions dominating the emission line component of the all-sky ISRF, plus an {\em in situ} WIM component behind the globules. Combining the line spectra of the bright rim and dark core regions allowed us to separate the scattered and {\em in situ} line components.      

After subtracting a model spectrum from the scattered light, there remains an excess emission between  $\sim 5500-8000$~\AA\ in DC~303 and more weakly in the TPN, which is reminiscent of the photoluminescence band detected in reflection nebulae and other objects and is called ERE. The nature of this excess emission in DC~303 remains inconclusive however.

\begin{acknowledgements}
We thank Chris Flynn and Laura Portinari for providing their detailed tables of stellar distributions and colours in the solar neighbourhood as derived from HIPPARCOS observations. 
We thank the referee, Adolf N.\ Witt, for his most constructive and helpful criticism.
This research was supported by the grants Nos.\ 1204415, 1201269, 122031 and 132291 of the Finnish Research Council for Science and Technology, as well as by grants of the Vilho, Yrj\"o and Kalle V\"ais\"al\"a Foundation of  The Finnish Academy of Science and Letters and the Magnus Ehrnrooth Foundation of The Finnish Society of Sciences and Letters. This publication makes use of data products from the Two Micron All Sky Survey, which is a joint project of the University of Massachusetts and the Infrared Processing and Analysis Center/California Institute of Technology, funded by the National Aeronautics and Space Administration and the National Science Foundation. This publication uses data from the Southern H-Alpha Sky Survey Atlas (SHASSA), which is supported by the National Science Foundation. The Wisconsin H-Alpha Mapper is funded by the National Science Foundation. This research has made use of SAOImage DS9, developed by Smithsonian Astrophysical Observatory. The Digitized Sky Surveys were produced at the Space Telescope Science Institute under U.S. Government grant NAG W-2166. The images of these surveys are based on photographic data obtained using the Oschin Schmidt Telescope on Palomar Mountain and the UK Schmidt Telescope. The plates were processed into the present compressed digital form with the permission of these institutions. The UK Schmidt Telescope was operated by the Royal Observatory Edinburgh, with funding from the UK Science and Engineering Research Council (later the UK Particle Physics and Astronomy Research Council), until 1988 June, and thereafter by the Anglo-Australian Observatory. The blue plates of the southern Sky Atlas and its Equatorial Extension (together known as the SERC-J), as well as the Equatorial Red (ER), and the Second Epoch [red] Survey (SES) were all taken with the UK Schmidt.
\end{acknowledgements}


\appendix

\section{Synthetic spectrum of the integrated starlight} 
Spectral synthesis is a common method in investigations of stellar  populations in external galaxies that are too distant to be resolved  into individual stars. In this section we address the opposite problem:  given the number densities and spatial distribution of the different types  of stars in the solar neighbourhood, what is the spectrum of the integrated starlight (ISL) in different directions of sky and for different  vantage points of an observer (or a globule) off the Galactic plane ? In addition to the ISL, the Galactic surface brightness also contains the diffusely scattered starlight component, called the diffuse Galactic light (DGL). It contributes ca.\ 
10--30\% of the total light and has to be explicitly considered when estimating the total intensity of Galactic light. Because its spectrum is a copy of the ISL spectrum, it does not influence the strengths of the spectral features (absorption lines, bands, discontinuities), which are the main topic of our comparison of the ISL with the observed  globule spectra. Our calculation of the ISL spectrum between 3000 and 10000~\AA\ follows the methods presented in \citet{m80a,m80b}. We here give a brief description of ISL spectral modelling appropriate for the comparison with the globule spectra. A more elaborate description and discussion of the results will be published in a separate paper. 

\subsection{Galaxy model}
A simple model of the Galactic structure is adopted in which the  stars and the dust are distributed in plane-parallel layers. The stars are divided into representative spectral groups and the number density of  stars per pc$^3$, $D_i(z)$, and the stellar emission  coefficient per pc$^3$, $j_i(z)$, in each group  $i$ are described by exponential density distributions in the 
$z-$direction:

\begin{equation}
D_i(z) = D_i(0)sech(z/h_{z,i})   
\end{equation}
\begin{equation}
j_i(z) = j_i(0)sech(z/h_{z,i}),          
\end{equation}

where  $D_i(0)$ and $j_i(0)$ are the densities in the galactic plane and  
$h_{z,i}$ the scale height. The $z$-dependence of the extinction coefficient  $a(z)$ (mag kpc$^{-1}$) is approximated with the formula

\begin{equation}
a(z) = a(0) exp(|z|/\beta_d),  
\end{equation}

where we adopted $a(0)=1.75$~mag/kpc (in V band) and $\beta_d$=134~pc. The extinction coefficient in the Galactic plane $a(0)$ is the average value for the distance range $d=0-2.5$~kpc  derived by \citet{sch67a,sch67b} from a statistical analysis of the colour excesses of 4700 stars near the Galactic plane ($|z|<75$~pc) as  provided by \citet{n66}. It agrees with the later estimates of e.g.\ \citet{h97} ($\approx 1.5$~mag/kpc) and \citet{mva98}. The scale height $\beta_d$ was adopted from \citet{mrr06} and is in accordance (within $\sim \pm 20$~pc) with other determinations for the solar neighbourhood within a few kpc.   

The results of \citet{n66}, which we used in our 1980 ISL modelling, have been derived for the more nearby Galactic   environment, $d \leq 500$~pc, and indicated an extinction coefficient  
$a(0)=2.46~ \mathrm{mag}\ \mathrm{kpc}^{-1}$ (V) and a scale height corresponding to $\beta_d$=40~pc. The ISL is, however, estimated to originate in a substantially higher volume around the Sun, reaching up to $d \geq 3$~kpc in the Galactic plane.
    
For the wavelength dependence of interstellar extinction the formula given
by \citet{c89} for $R_V=3.1$ was adopted. 

The clumpiness of the dust  distribution has a profound effect on the ISL intensity. For a given mean extinction coefficient the mean ISL intensity for the clumpy dust distribution is higher than that for the homogeneous model. For a model consisting of a uniform distribution of stars and dust clouds \citet{cm50} derived the following expression for the mean ISL intensity:

\begin{equation}
I = \frac{j}{\nu}\frac{1}{1-q}[1 - e^{-\bar{n}(1-q)}].   
\end{equation}

Here $\nu$ is the average number of clouds per kpc, $\bar{n}$ the mean  number of clouds along the whole line of sight and $q$ the average transmissivity of a dust cloud $q = \langle exp(-\tau)\rangle$, where $\tau$  is the optical depth of one cloud. The parameters for the dust clumpiness correction were adopted according to  the two-component model of \citet{sch67a,sch67b}: there are two kinds of clouds with a mean optical thickness of 
$\bar{\tau_1}(V)=0.24$ and $\bar{\tau_2}(V)=1.5$ and a relative fequency of 
$\nu_1 = 0.9$ and $\nu_2 = 0.1$.

The numerical integration along each line of sight was performed in steps  of 10~pc and extended to $z_{max}$=3600~pc, which is well outside the dust distribution layer in the directions $|b|\ne 0$. For distances $z \geq 3600$~pc the ISL contribution was estimated and found to be negligible. For $b=0$~deg the maximun distance along the plane was set to 10~kpc because high foreground  extinction (17.5~mag at $z=0$, 8.3~mag at $z=100$) prevents any  significant starlight to penetrate from beyond  this distance. For directions close to the Galactic plane, $b \leq 10$~deg, the maximum distance to which the numerical integration was extended from the plane  was limited to 
$z_{max}<$3600~pc by the condition not to exceed the maximun distance of 10~kpc along the plane. For details of the model and the other formulae used  see \citet{m80a}.

\subsection{Stellar distribution parameters}

The division of stars into representative spectral groups covering the different parts of the HR-diagram was made according to the  approach of 
\citet{f06} based on the analysis of the {\it Hipparcos} data base. Chris Flynn and Laura Portinari kindly provided us with their results in detailed tabular form. In these tables the stars are divided into the the  following seven categories: {\it (i, ii)} main sequence (thin and thick disk),  
{\it (iii, iv)} clump stars (thin and thick disk),  {\it (v, vi)} old giants  (thin and thick disk), and {\it (vii)} young giants. For each category the tables give as a function of $M_V$ the stellar number density  in the Galactic plane, $D(0)$, and the scale height $h_z$ for a  distribution of the form 
$D(z)=D(0)$sech$(z/h_z)$. In addition, the colour indices $B-V$ and $V-I_c$ based on the {\it Hipparcos} and {\it Tycho} catalogues were derived for each magnitude slot of the seven categories. Because of the limited distance range of {\it Hipparcos}, its coverage for the supergiants was sparse. We complemented this part by using the compilation of \citet{w92}. Our adopted parameters for the stellar distributions are shown in  
Table~\ref{tab:StellarDistribution}.

\subsection{Stellar spectral library}

For the comparison with the 3500--9000~\AA\ scattered light spectra of  the globules we need a spectral library with good spectral type and wide  wavelength coverage but with only a relatively modest  spectral resolution. The classical \citet{p98} library fulfils these  conditions best. In order to choose the best template stars from the library for each of our stellar slots as given in Table~\ref{tab:StellarDistribution} we used both the absolute magnitudes $M_V$ and the colour indices $B-V$ and $V-I_c$ as selection criteria. In most cases a satisfactory match with a Pickles  template spectrum was possible. In a few cases we adopted the average of two  nearby spectral templates. The Pickles catalogue numbers and  spectral types for the chosen stellar templates are given in Table~\ref{tab:StellarDistribution}.

\subsection{Results}

\begin{figure*}[htbp]
     \centering
     \begin{minipage}[t]{0.5\hsize}
       \centering
       \includegraphics[width=0.85\linewidth]{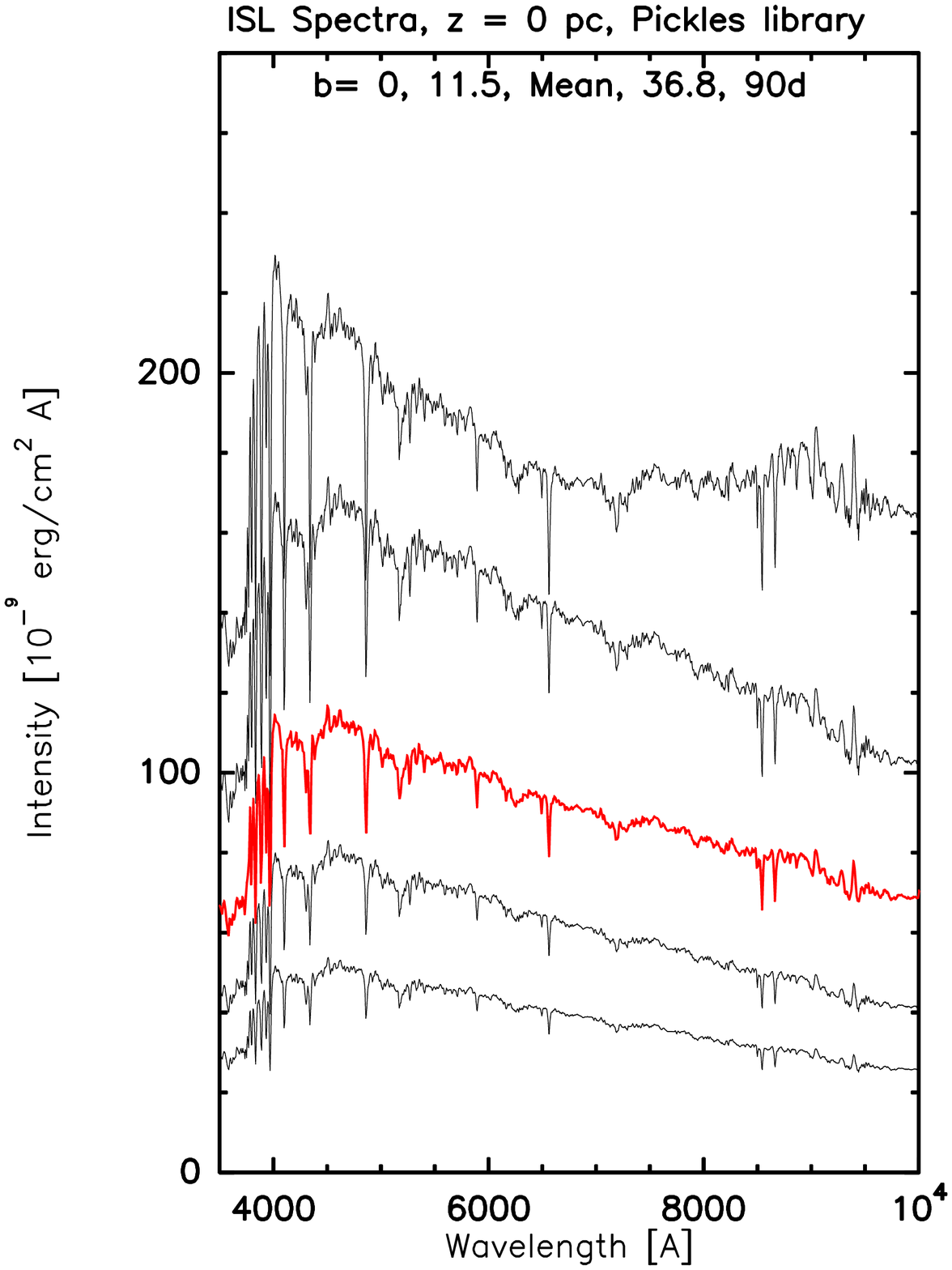}
     \end{minipage}%
     \begin{minipage}[t]{0.5\hsize}
        \centering
        \includegraphics[width=0.85\linewidth]{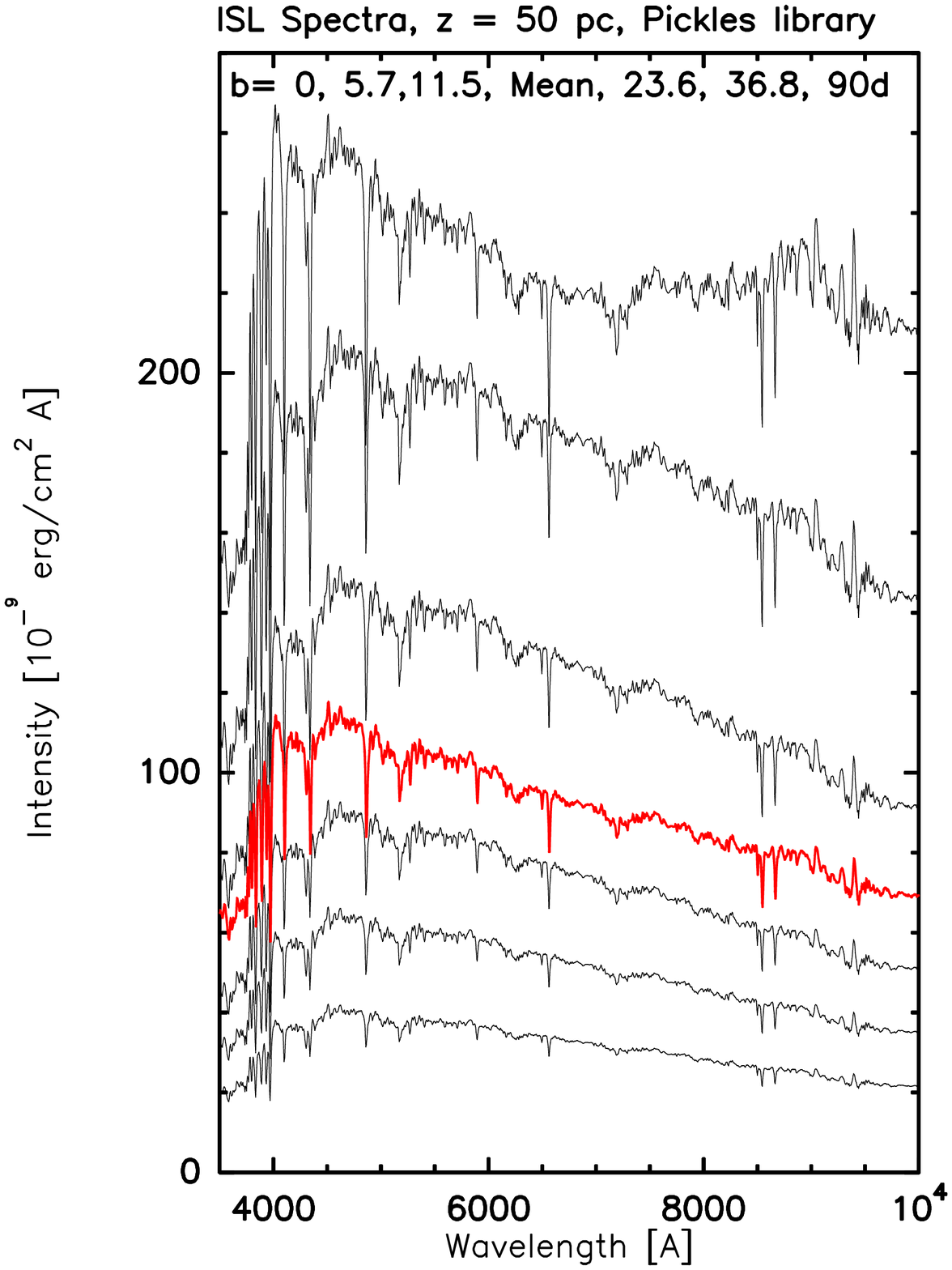}
     \end{minipage}\\[20pt]
     \begin{minipage}[t]{0.5\hsize}
       \centering
       \includegraphics[width=0.85\linewidth]{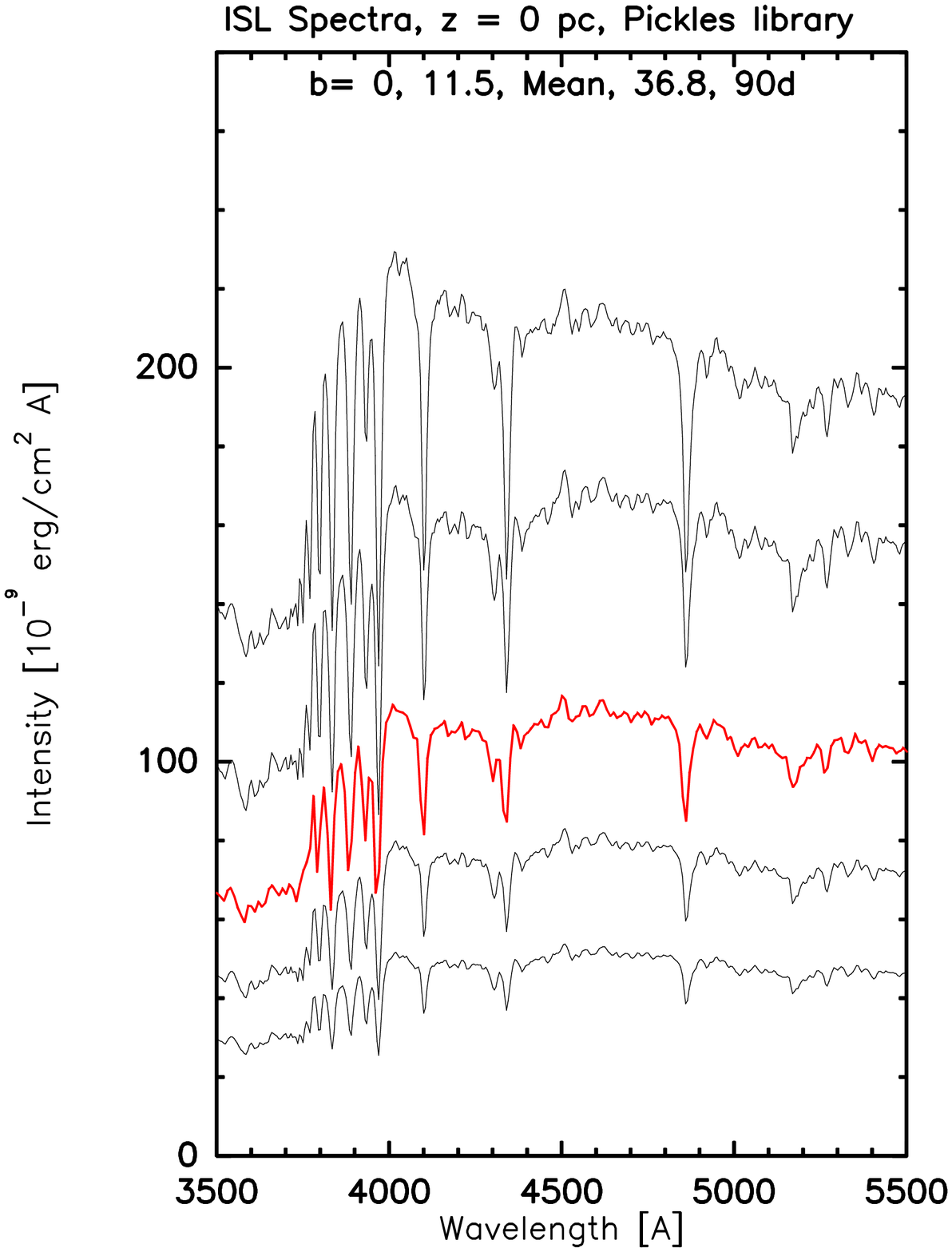}
     \end{minipage}%
     \begin{minipage}[t]{0.5\hsize}
       \centering
       \includegraphics[width=0.85\linewidth]{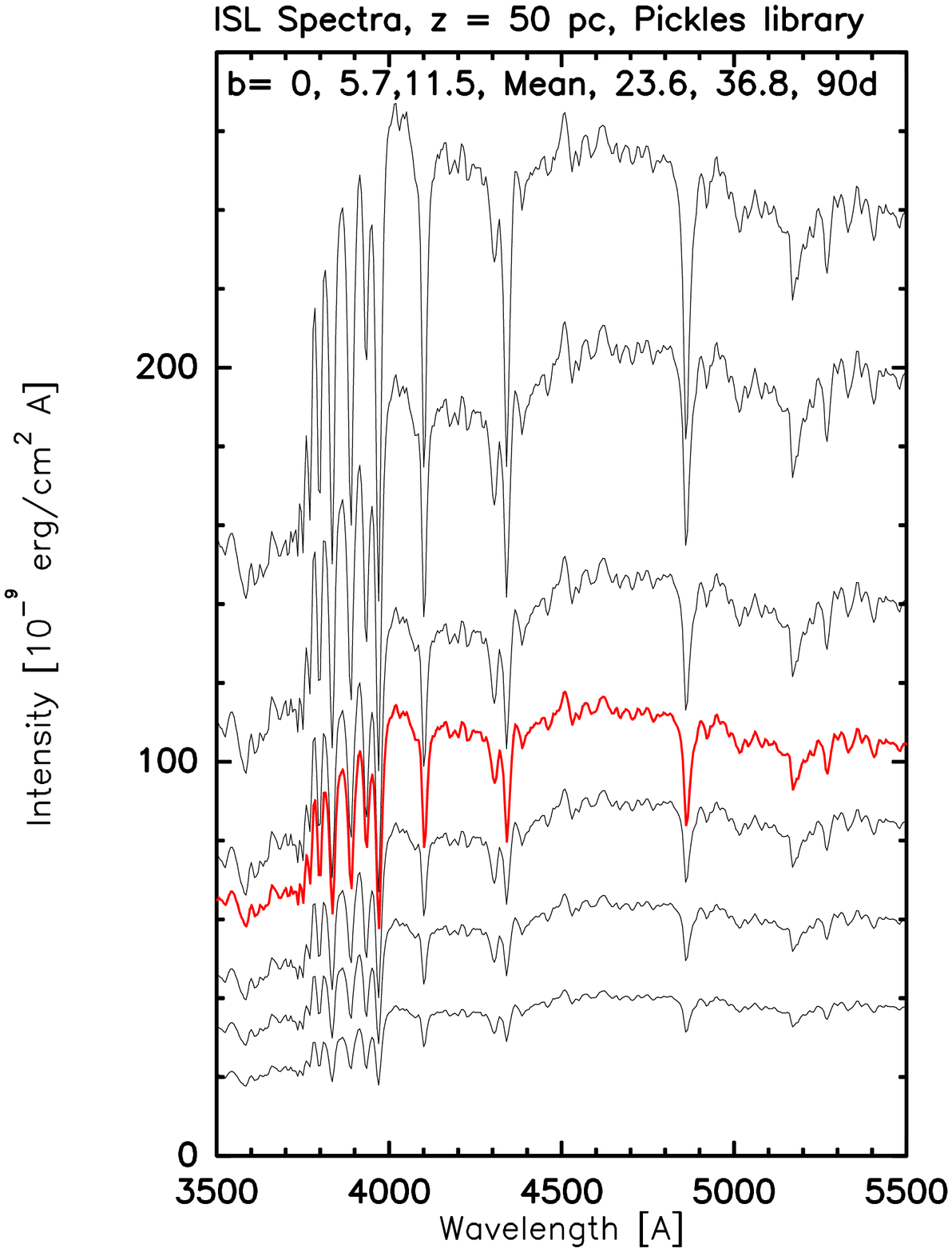}
     \end{minipage}%
      \caption{Integrated starlight spectra for different Galactic latitudes and for two positions of the observer, $z=0$ and  $z=50$~pc. The upper panels show the wavelength range 3500--10000~\AA, the lower panels show the wavelength range 3500--5500~\AA. For $z=0$~pc the spectra are shown from bottom to top for $b = 90, 36.8, 11.5$, and 0~deg as a black line. For 
$z=50$~pc the spectra are shown from bottom to top for 
$b = 90, 36.8, 23.6, 11.5, 5.7$, and 0~deg as a black line.  The mean spectra over the sky are shown as a red line. All spectra have the same scale and zero point.} 
\label{fig:ISLspectra} 
\end{figure*}

We show in Fig.~\ref{fig:ISLspectra} the resulting ISL spectra for several galactic latitudes from $|b|$=0 to 90~deg for an observer located in the plane ($z=0$~pc).  In Fig.~\ref{fig:ISLspectra} we also show ISL spectra for 
$z=50$~pc, corresponding to the  estimated distance of the two globules.   The two spectra at $b$=11.5 and 23.8~deg are bracketing the latitude range of the two globules, b=-14.2~deg for DC~303 and b=-15.9~deg for the TPN.  We also show the mean spectrum over the whole sky for the two  $z$ values. It can be seen that the absorption lines, bands, and discontinuities are very similar in all these ISL spectra. The ISL intensity for intermediate and high positive  Galactic latitudes decreases, as expected, when the observer moves  to 
$z=50$~pc. On the opposite side of the sky ($b<0$), not shown in the  figure, the ISL intensity increases correspondingly.  The ISL intensity at Galactic equator increases, because the extinction coefficient  at $z=50$~pc is only 
69\% of its value at  $z=0$. The mean ISL intensity  is seen to decrease modestly between  $z=0$ and $z=50$~pc.

\onecolumn
\begin{longtab}
\begin{longtable}{rrrrcrcrrlr}
 \caption{Data for stellar distribution parameters} \\
 \hline\hline
 No. & $M_V$ & \multicolumn{2}{c}{$M_V$ range} & $D_0$ & $h_z$ & $j_0$ & 
 $B-V$ & $V-I$ & Pickles   & Pickles\\
 &    &     &      &     &    &    &     &    & Sp Class & No.   \\
 & mag   & \multicolumn{2}{c}{mag}  & stars/pc$^3$ &  pc & 
 $M_V=0$ stars/pc$^3$ &  mag &   mag &  & \\
\hline
\endfirsthead
\caption{continued.}\\
\hline\hline
 No. & $M_V$ & \multicolumn{2}{c}{$M_V$ range} & $D_0$ & $h_z$ & $j_0$ & 
 $B-V$ & $V-I$ & Pickles & Pickles\\
 &    &   &      &     &    &      &     &    & Sp Class & No.   \\
 & mag   & \multicolumn{2}{c}{mag}  & stars/pc$^3$ &  pc & 
 $M_V=0$ stars/pc$^3$ &  mag &   mag &  & \\
\hline
\endhead
\hline
\endfoot
  &     &     &       &      &      &      &      &    &     &    \\
\multicolumn{11}{l}{Main sequence, thin disk}\\
 1   &    -4.0 & -4.20 & -3.25 &0.46E-06 &  56. & 0.13E-04 &  -0.28 & -0.28& B0/B1 V &  3,4\\
 2   &    -2.45& -3.20 &-1.85  &0.34E-05 &  56. & 0.32E-04 &  -0.23 & -0.23& B1    V &  4 \\
 3   &    -1.2 & -1.80 &-0.75  &0.13E-04 &  56. & 0.38E-04 &  -0.13 & -0.16& B5-7  V &  6\\
 4   &    -0.25& -0.70 & 0.20  &0.44E-04 &  57. & 0.53E-04 &  -0.10 & -0.10& B8    V &  7 \\
 5   &     0.65&  0.25 & 0.95  &0.92E-04 &  68. & 0.52E-04 &   0.01 &  0.01& A0    V &  9 \\
 6   &     1.3 &  1.00 & 1.60  &0.15E-03 &  82. & 0.44E-04 &   0.14 &  0.14& A5    V & 12\\
 7   &     1.8 &  1.65 & 1.95  &0.11E-03 &  92. & 0.21E-04 &   0.26 &  0.26& A7    V & 13\\
 8   &     2.1 &  2.00 & 2.30  &0.14E-03 & 107. & 0.20E-04 &   0.32 &  0.32& A7/F0 V &  13,14\\
 9   &     2.7 &  2.35 & 2.90  &0.41E-03 & 140. & 0.36E-04 &   0.42 &  0.42& F0    V &  14 \\
10   &     3.1 &  2.95 & 3.30  &0.42E-03 & 176. & 0.24E-04 &   0.49 &  0.49& F5    V &  16 \\
11   &     3.5 &  3.35 & 3.75  &0.59E-03 & 206. & 0.23E-04 &   0.55 &  0.56& F6    V &  18\\
12   &     4.0 &  3.80 & 4.20  &0.71E-03 & 220. & 0.18E-04 &   0.59 &  0.62& F8    V &  20\\
13   &     4.4 &  4.25 & 4.55  &0.65E-03 & 220. & 0.11E-04 &   0.61 &  0.66& G0    V &  23\\
14   &     4.7 &  4.60 & 4.90  &0.74E-03 & 220. & 0.93E-05 &   0.64 &  0.69& G2    V &  26\\
15   &     5.1 &  4.95 & 5.30  &0.92E-03 & 220. & 0.82E-05 &   0.70 &  0.73& G5    V &  27\\
16   &     5.5 &  5.35 & 5.70  &0.94E-03 & 220. & 0.58E-05 &   0.78 &  0.81& G8    V &  30\\
17   &     5.9 &  5.75 & 6.15  &0.11E-02 & 220. & 0.45E-05 &   0.88   &0.91& K0    V &  31\\
18   &     6.4 &  6.20 & 6.90  &0.19E-02 & 220. & 0.47E-05 &   0.94   &0.99& K2    V &  33\\
19   &     7.35&  6.95 & 8.05  &0.33E-02 & 220. & 0.35E-05 &   1.10   &1.27& K4    V &  35\\
20   &     8.8 &  8.10 & 9.35  &0.46E-02 & 220. & 0.15E-05 &   1.42   &1.77& M0    V &  38\\
21   &     9.9 &  9.40 &11.10  &0.11E-01 & 220. & 0.94E-06 &   1.50   &2.17& M2.5  V &  41\\
22   &    12.2 & 11.15 &12.95  &0.15E-01 & 220. & 0.27E-06 &   1.62   &2.64& M4    V &  43\\
  &     &     &       &      &      &      &      &    &     &    \\
\multicolumn{11}{l}{Main sequence, thick disk}\\
23   &     2.1 &  2.00 & 2.30  &0.39E-06 & 670. & 0.50E-07 &   0.27 &  0.27& A7    V &  13\\
24   &     2.7 &  2.35 & 2.90  &0.45E-05 & 670. & 0.38E-06 &   0.37 &  0.37& F0    V &  14\\
25   &     3.1 &  2.95 & 3.30  &0.80E-05 & 670. & 0.44E-06 &   0.44 &  0.44& F2    V &  15\\
26   &     3.5 &  3.35 & 3.75  &0.50E-04 & 670. & 0.18E-05 &   0.50 &  0.51& wF5   V &  17\\
27   &     4.0 &  3.80 & 4.20  &0.78E-04 & 670. & 0.20E-05 &   0.54 &  0.57& wF8   V &  21\\
28   &     4.4 &  4.25 & 4.55  &0.72E-04 & 670. & 0.13E-05 &   0.56 &  0.61& F8    V &  20\\
29   &     4.7 &  4.60 & 4.90  &0.82E-04 & 670. & 0.10E-05 &   0.59 &  0.63& G0    V &  23\\
30   &     5.1 &  4.95 & 5.30  &0.10E-03 & 670. & 0.91E-06 &   0.65 &  0.68& wG5   V &  28\\
31   &     5.5 &  5.35 & 5.70  &0.10E-03 & 670. & 0.65E-06 &   0.73 &  0.76& G8    V &  30\\
32   &     5.9 &  5.75 & 6.15  &0.12E-03 & 670. & 0.50E-06 &   0.83 &  0.86& K0    V &  31\\
33   &     6.4 &  6.20 & 6.90  &0.21E-03 & 670. & 0.52E-06 &   0.89 &  0.94& K2    V &  33\\
34   &     7.35&  6.95 &8.05   &0.37E-03 & 670. & 0.38E-06 &   1.06 &  1.22& K4    V &  35\\
35   &     8.8 &  8.10 & 9.35  &0.51E-03 & 670. & 0.17E-06 &   1.37 &  1.72& M0    V &  38\\
36   &     9.9 &  9.40 &11.10  &0.13E-02 & 670. & 0.10E-06 &   1.45 &  2.12& M2/2.5 V & 40,41\\
37   &    12.2 & 11.15 &12.95  &0.17E-02 & 670. & 0.30E-07 &   1.56 &  2.59& M3/M4 V &  42,43\\
  &     &     &       &      &      &      &      &    &     &    \\
\multicolumn{11}{l}{Clump stars, thin disk}\\
38   &    -1.6 & -2.10 & -1.15 &0.36E-06 & 180. & 0.13E-05 &   2.1  &  2.1 & M3    III & 98,113\\
39   &    -1.0 & -1.10 & -0.95 &0.25E-06 & 180. & 0.63E-06 &   1.77 &  1.76& M1    III & 96\\
40   &    -0.85& -0.90 & -0.80 &0.22E-06 & 180. & 0.49E-06 &   1.57 &  1.60& K5/M0 III & 93,95\\
41   &    -0.7 & -0.75 &-0.60  &0.34E-06 & 180. & 0.64E-06 &   1.45 &  1.45& K5    III & 93 \\
42   &    -0.5 & -0.55 &-0.45  &0.27E-06 & 180. & 0.43E-06 &   1.26 &  1.26& wK4   III & 91 \\
43   &    -0.35& -0.40 &-0.30  &0.27E-06 & 180. & 0.37E-06 &   1.16 &  1.14& wK3   III & 88 \\
44   &    -0.2 & -0.25 &-0.10  &0.52E-06 & 180. & 0.61E-06 &   1.02 &  1.02& K2    III & 85\\
45   &     0.3 & -0.05 & 0.65  &0.29E-04 & 180. & 0.18E-04 &   0.88 &  0.89& G5II/wG8III & 110,77\\
46   &     0.95&  0.70 & 1.20  &0.52E-04 & 180. & 0.25E-04 &   1.05 &  1.05& wK2   III & 85\\
  &   &    &     &     &      &     &      &    &      &    \\
\multicolumn{11}{l}{Clump stars, thick disk}\\
47   &    -0.7 & -0.75 &-0.60  &0.44E-08 & 670. & 0.78E-08 &   1.45 &  1.45& K5    III & 93\\
48   &    -0.5 & -0.55 &-0.45  &0.60E-07 & 670. & 0.94E-07 &   1.26 &  1.26& wK4   III & 91\\
49   &    -0.35& -0.40 &-0.30  &0.18E-06 & 670. & 0.25E-06 &   1.15 &  1.14& wK3   III & 88\\
50   &     0.4 & -0.25 & 1.00  &0.12E-04 & 670. & 0.65E-05 &   0.90 &  0.95& wG8III/G5II& 77,110\\
  &    &     &     &      &      &      &     &    &     &    \\
\multicolumn{11}{l}{Old giants, thin disk}\\
51   &    -0.6 & -1.25 & 0.00  &0.18E-04 & 180. & 0.29E-04 &   1.45 &  1.45& K5    III & 93\\
52   &     0.5 &  0.05 & 1.00  &0.24E-04 & 180. & 0.16E-04 &   1.31 &  1.28& wK4   III & 91\\
53   &     1.5 &  1.05 & 2.00  &0.50E-04 & 180. & 0.12E-04 &   1.07 &  1.06& K1    III & 81\\
54   &     2.5 &  2.05 & 3.00  &0.67E-04 & 180. & 0.65E-05 &   0.95 &  0.95& K0    IV &  57 \\
55   &     3.5 &  3.05 & 4.00  &0.13E-03 & 180. & 0.49E-05 &   0.86 &  0.86& G8    IV &  56\\
56   &     4.35&  4.05 & 4.65  &0.64E-05 & 180. & 0.15E-06 &   0.85 &  0.85& G8    IV &  56\\
   &    &       &       &      &      &       &      &    &       &    \\
\multicolumn{11}{l}{Old giants, thick disk}\\
57   &    -1.0 & -2.00 & 0.00  &0.21E-05 & 670. & 0.50E-05 &   1.45 &  1.45& K5   III &  93\\
58   &     0.5 &  0.05 & 1.00  &0.24E-05 & 670. & 0.14E-05 &   1.28 &  1.28& K4   III &  90\\
59   &     1.5 &  1.05 & 2.00  &0.65E-05 & 670. & 0.15E-05 &   1.07 &  1.06& wK1  III &  82\\
60   &     2.5 &  2.05 & 3.00  &0.15E-04 & 670. & 0.15E-05 &   0.95 &  0.95& K0   IV &   57\\
61   &     3.25&  3.05 & 3.45  &0.99E-05 & 670. & 0.49E-06 &   0.88 &  0.88& G8   IV &   56\\
  &    &     &       &     &      &      &      &    &       &    \\
\multicolumn{11}{l}{Young giants, thin disk \tablefootmark{a}} \\
62   &    -0.5 & -1.00 & 0.00  &0.60E-05 &  56. & 0.10E-04 &   1.30 &  1.44& K4   III &   90  \\
63   &     0.5 &  0.05 & 0.95  &0.57E-05 &  56. & 0.38E-05 &   0.87 &  1.28& K3   III &   87  \\
64   &     1.7 &  1.00 & 2.45  &0.84E-05 & 100. & 0.17E-05 &    -   &  1.03& K1   III &   81  \\
65 \tablefootmark{b}  & -5.14 &-&- & 0.20E-06 & 56.&  2.28E-05 & -&  -  & 
O5/B5 V &  1,117  \\           
66  \tablefootmark{c} & -6.05 &-&- & 0.33E-07 & 56. & 0.84E-05&- & -  & 
A2/G0/G2 I & 120, 122, 125  \\   
67 \tablefootmark{d} & -6.34 &-&- & 0.32E-07 & 56. & 1.10E-05& -& -  & 
K2/M2 I & 128,131 \\        
68 \tablefootmark{e} & -5.79 &- &- & 0.13E-07 & 56. & 0.27E-05 & -& -   & 
M3 II   & 113    \\         
\end{longtable}
\tablefoot{
\tablefoottext{a}{Parameters for supergiants are according to Wainscoat et~al.\ (1992)}  
\tablefoottext{b}{OB}
\tablefoottext{c}{A-G I-II}
\tablefoottext{d}{K-M2 I-II}
\tablefoottext{e}{M3-M4 I-II} }
\label{tab:StellarDistribution}
\end{longtab}
\twocolumn

\end{document}